\documentclass[twocolumn]{aastex631}
\usepackage[T1]{fontenc}
\usepackage[utf8]{inputenc}
\usepackage{CJKutf8}
\usepackage{xcolor}
\usepackage{graphicx}
\usepackage{listings}
\usepackage{amsmath}
\usepackage{comment}
\usepackage{tabularx}
\usepackage{booktabs}
\definecolor{mauve}{rgb}{0.88, 0.69, 1.0}
\definecolor{oldmauve}{rgb}{0.4, 0.19, 0.28}
\defcitealias{2023arXiv230202611A}{A23}

\newcommand{\teff}{$T_{\rm eff}$}
\newcommand{\logg}{$\log ~ g$}
\newcommand{\moh}{[M/H]}
\newcommand{\aom}{[$\alpha$/M]}
\newcommand{\emoh}{$\sigma_{\rm {[M/H]}}$}
\newcommand{\eteff}{$\sigma_{T_{\rm eff}}$}
\newcommand{\elogg}{$\sigma_{\log ~ g}$}
\newcommand{\eaom}{$\sigma_{[\alpha/{\rm M}]}$}
\newcommand{\gaia}{\textsl{Gaia}}
\newcommand{\apogee}{\textsl{APOGEE}}
\newcommand{\xp}{\textsl{AspGap}}

\definecolor{rb4}{HTML}{27408B}

\usepackage{amsmath}

\begin{document}

%%%%%%%%%%%%%%%%%%%%%%%%%%%%%%%%%%%%%%%%%%%%%%%%%%%%%%%
%
%
%
%                 Title & Authors
%
%
%
%%%%%%%%%%%%%%%%%%%%%%%%%%%%%%%%%%%%%%%%%%%%%%%%%%%%%%%%%
\title{
    \xp: Augmented Stellar Parameters and Abundances for 23 million RGB stars from \textsl{Gaia} XP low-resolution spectra }

\author[0000-0002-3651-5482]{Jiadong Li
(\begin{CJK*}{UTF8}{gbsn}李佳东\ignorespacesafterend\end{CJK*})
}

\affiliation{Key Lab of Space Astronomy and Technology, National Astronomical Observatories, Beijing, 100101, China}
\affiliation{Center for Cosmology and Particle Physics, Department of Physics, New York University, 726 Broadway, New York, NY 10003, USA}
\affiliation{University of Chinese Academy of Sciences, Beijing, 100049, China}
\affiliation{Max-Planck-Institut für Astronomie, Königstuhl 17, D-69117 Heidelberg, Germany}

\author[0000-0001-8432-7788]{Kaze W. K. Wong}
\affiliation{Center for Computational Astrophysics, Flatiron Institute, 162 Fifth Avenue, New York, NY 10010, USA}

\author[0000-0003-2866-9403]{David W. Hogg}
\affiliation{Center for Cosmology and Particle Physics, Department of Physics, New York University, 726 Broadway, New York, NY 10003, USA}
\affiliation{Max-Planck-Institut für Astronomie, Königstuhl 17, D-69117 Heidelberg, Germany}
\affiliation{Center for Computational Astrophysics, Flatiron Institute, 162 Fifth Avenue, New York, NY 10010, USA}

\author[0000-0003-4996-9069]{Hans-Walter Rix}
\affiliation{Max-Planck-Institut für Astronomie, Königstuhl 17, D-69117 Heidelberg, Germany}

\author[0000-0002-0572-8012]{Vedant Chandra}
\affiliation{Center for Astrophysics $\mid$ Harvard \& Smithsonian, 60 Garden St, Cambridge, MA 02138, USA}

\begin{abstract}
\noindent
We present \xp, a new approach to infer stellar labels from low-resolution \textsl{Gaia} XP spectra, including precise [$\alpha$/M] estimates for the first time. 
\xp\ is a neural-network based regression model trained on APOGEE spectra. 
In the training step, \xp\ learns to use XP spectra not only to predict stellar labels but also the high-resolution APOGEE spectra that lead to the same stellar labels. 
The inclusion of this last model component  --- dubbed the \textit{hallucinator} --- creates a more physically motivated mapping and significantly improves the prediction of stellar labels in the validation, particularly of [$\alpha$/M]. 
For giant stars, we find cross-validated \textit{rms} accuracies for $T_{\rm eff}$, log~$g$, [M/H], [$\alpha$/M] of $\sim 1\%$, 0.12\,dex, 0.07\,dex, 0.03\,dex, respectively. 
We also validate our labels through comparison with external datasets and through a range of astrophysical tests that demonstrate that we are indeed determining [$\alpha$/M] from the XP spectra, rather than just inferring it indirectly from correlations with other labels. 
We publicly release the \xp{} codebase, along with our stellar parameter catalog for all giants observed by Gaia~XP. 
\xp\ enables new insights into the formation and chemo-dynamics of our Galaxy by providing precise [$\alpha$/M] estimates for 23 million giant stars, including 12 million with radial velocities from \textsl{Gaia}.

%A key part of \textsl{Gaia} DR3 is the BP/RP (XP) spectroscopic sample, which provides 220 million low-resolution spectra. These comprise an unprecedented homogeneous data set for understanding the abundances and dynamics of the Milky Way: The large luminosities of red-giant-branch (RGB) stars make them key to comprehensive studies of the Milky Way. Here we introduce \xp, a novel regression model for inferring stellar parameters from the XP spectra, augmented by transfer learning from high-resolution \textsl{APOGEE} spectra. Using a training set of stellar labels from \textsl{APOGEE}, we demonstrate that \xp, can deliver effective temperature, surface gravity, metallicity, and $\alpha$ abundance to accuracies of $\sim$ 1\,percent, 0.12\,dex, 0.07\,dex, and 0.03\,dex, respectively.The $\alpha$ abundance is validated in both \gaia-Enceladus/Sausage and the Large Magellanic Cloud, providing further confidence in its accuracy. We use \xp\ to deliver a stellar parameter catalog with fundamental parameters, including the overall metal abundance [M/H], as well as $\alpha$-abundance for 24 million RGB stars. We release a catalog of \xp\ parameters and open-source code that can be extended to the estimation of other stellar parameters.
\end{abstract}

%% Keywo rds should appear after the \end{abstract} command. 
% \keywords{}

%              Section: Introduction
%
%
%
%%%%%%%%%%%%%%%%%%%%%%%%%%%%%%%%%%%%%%%%%%%%%%%%%%%%%%%%%
\section{Introduction} \label{sec:intro}
% \noindent
% The study of the star formation and Galactic enrichment history of our Galaxy is a crucial approach to understanding the nature of galaxy evolution and cosmology. 
% The investigation of star formation and Galactic enrichment history plays a vital role in comprehending the evolution of galaxies and cosmology. 
Understanding the star formation and Galactic enrichment history of the Milky way is essential for gaining insights into the broader context of galaxy evolution and cosmology. 
For decades, the field of \textit{Galactic Archaeology} has been dedicated to unraveling the formation history of our own Galaxy. This endeavor has been greatly aided by large-scale spectroscopic surveys such as the Sloan Digital Sky Survey (SDSS)/The Apache Point Observatory Galactic Evolution Experiment (\apogee) \citep{Majewski2017}, the Milky Way Mapper of SDSS-V \citep{Kollmeier2017, Almeida2023}, the Large Sky Area Multi-Object Fiber Spectroscopic Telescope (LAMOST) \citep{Cui2012, Deng2012, Luo2012}, GALactic Archaeology with HERMES (GALAH) \citep{Buder2021}, and the 4-metre Multi-Object Spectroscopic Telescope (4MOST) \citep{deJong2019}. 
These surveys have played a crucial role in providing extensive spectroscopic data and enabling detailed investigations into the chemical and dynamical properties of stars in the Milky Way.
% As large spectroscopic surveys, e.g., Sloan Digital Sky Survey (SDSS)/ The Apache Point Observatory Galactic Evolution Experiment (\apogee) \citep{Majewski2017}, LAMOST \citep{Cui2012, Deng2012, Luo2012} proceed, our understanding of {\it Galactic Archaeology} has dramatically advanced due to the availability of large datasets with known fundamental properties of stars in the Milky Way (MW). 
In addition to the spectroscopic data, the European Space Agency's (ESA) \gaia\ mission \citep{2016A&A...595A...1G} has played a pivotal role in this field by observing billions of stars with unprecedented precision. 
The \gaia\ mission has provided measurements of parallax and proper motion, enabling the construction of a comprehensive six-dimensional phase-space information  of our Galaxy, revolutionizing our knowledge of the structure of the Milky Way (MW).
By combining astrometric information, ongoing and future spectroscopic surveys have the potential to significantly expand our understanding of fundamental galactic astronomy. 
These surveys can broaden the distribution range of atmospheric parameters and chemical abundances, leading to valuable insights into various aspects such as the formation history of the Milky Way \citep{xiang2022b}, 
the variation in the stellar initial mass function across different chemical environments and star formation histories \citep{jdli2023},
and the discovery of the existence of very massive stars in the early universe \citep{xing2023metal}.

In the recent \gaia\ Data Release 3 (DR3) \citep{GaiaCollaboration2022}, a substantial number of approximately 220 million low-resolution spectra of stars have been made available \citep{GaiaCollaboration2022, DeAngeli2022, Montegriffo2022}. 
These spectra, obtained through the combined observations of the blue photometer (BP) and red photometer (RP) in the \gaia\ mission, provide essential information on stellar parameters, including chemical abundance measurements. 
The wavelength range covered by the combined BP/RP (XP) spectra spans from 3,300 to 10,500 $\mathrm{\AA}$, with a resolution $\mathcal{R}\approx 15-85$ \citep{Andrae2022}. These spectra serve as valuable resources for inferring stellar parameters, distances, and extinctions for stars within the Milky Way \citep{Andrae2022}. 

The potential of using the \gaia\ XP spectra for stellar parameter estimation was initially explored by \cite{Liu2012}. 
Subsequently, the \gaia\ General Stellar Parameterizer from Photometry, known as {\tt\string GSP-phot} \citep{Andrae2022}, applied a Bayesian forward-modeling approach to fit the XP spectra and produced a homogeneous catalog containing effective temperature (\teff), surface gravity, and metallicity estimates for approximately 471 million stars with $G$-band magnitudes brighter than 19. 
Although {\tt\string GSP-phot} provides valuable information on fundamental stellar parameters, the estimation of metal abundance (\moh) is not without its limitations due to the characteristics of the \gaia\ XP system and the limited information it provides about \moh~\citep{Andrae2022,2023arXiv230202611A}.
A theoretical study by \cite{Ting2017a} has demonstrated that valuable information on element abundances, including [Fe/H] and $\alpha$--abundance (\aom), can be gleaned from low-resolution spectra, although stellar labels become degenerate at $\mathcal{R}\leq 100$.
Remarkably, the precision of these element abundances remains largely unaffected by the resolution of the spectra as long as the exposure time and number of detector pixels are held constant. 
In fact, low-resolution spectra such as \gaia\ XP offer the advantage of higher signal-to-noise ratio (S/N) per pixel and a broader wavelength coverage within a single observation. 
These characteristics highlight the potential of low-resolution spectra to deliver accurate measurements of element abundances.

Accurately determining chemical abundances from XP spectra using traditional model-driven methods, which rely on comparing observed spectra to stellar spectral libraries, presents challenges due to the inherent systematic differences in flux calibration between observations and synthetic spectra.
Theoretical spectra and observed spectra often exhibit separate distributions in the high-dimensional flux space because of the imperfections in the theoretical spectra and errors introduced by observation conditions and instrument effects \citep{Wang2023b}. 
This calibration issue becomes particularly problematic when attempting to identify $\alpha$-sensitive spectral lines \citep{Gavel2021}.
Given the advantages of low-resolution spectra and the challenges associated with traditional model-driven methods, employing a data-driven approach for estimating chemical abundances from low-resolution spectra becomes a natural and promising choice \citep{2017ApJ...843...32T}. 
By leveraging the information contained in the data itself, data-driven methods can overcome the limitations of model-driven approaches and provide more robust and accurate estimates of chemical abundances.

Data-driven methods and machine learning techniques have become widely adopted for deriving stellar labels (parameters) from large volumes of low-resolution spectra. 
These methods offer an alternative approach to traditional model-driven methods and have shown great success in accurately estimating stellar properties in recent years (e.g., \citealt{Ness2015, 2017ApJ...843...32T, Ting2019a, zhang2020a, Xiang2022}).

There are two main categories of data-driven methods: empirical forward models and discriminative models. 
In the empirical forward model approach, models are built to predict the spectrum based on the stellar parameters \citep{Ness2015, Casey2016, Ho2017, 2017ApJ...843...32T, Ting2019a, zhang2020a, 2021ApJS..253...45L, Xiang2022, 2023arXiv230303420Z}. 
These models utilize a large training dataset with known stellar parameters to establish the relationship between the spectra and the stellar labels. 
By applying these models to new spectra, the stellar parameters can be inferred.
On the other hand, discriminative models take spectra as input and output stellar labels \citep{Leung2019, Rix2022, 2023arXiv230202611A, 2023arXiv230317676Y}. 
These models are trained using a labeled dataset, where both the spectra and the corresponding stellar parameters are known. 
The models learn the complex mapping between the input spectra and the desired output labels, allowing them to predict stellar parameters for unseen spectra.
Both empirical forward models and discriminative models have their strengths and applications. 
Empirical forward models directly predict the spectra based on the stellar parameters, which can be useful for studying the physical processes shaping the spectra and for generating synthetic spectra for stellar population synthesis. 
The discriminative models, on the other hand, provide a more direct approach to infer stellar labels from spectra, which is beneficial when the focus is on estimating stellar parameters for large datasets efficiently.
Recently, \cite{Leung2023} demonstrated that a single Transformer-based neural network trained on heterogeneous spectroscopic and photometric datasets could perform both discriminative tasks like predicting stellar parameters from spectra as well as generative tasks such as generating spectra from parameters, which opens up some new ideas on how to use spectra and stellar labels.

\textsl{The Cannon} \citep{Ness2015} is a data-driven generative model that was introduced for spectroscopic data analysis.
This approach involves establishing mappings between known stellar labels and spectra using a training dataset. 
Once the data-driven model is trained, it can be applied to infer labels for observed spectra. 
\textsl{The Cannon} has demonstrated its effectiveness in deriving labels for high-resolution \apogee\ spectra \citep{Ness2015, Casey2016}, as well as low-resolution spectra from surveys like LAMOST \citep{Ho2017}. 
Data-driven methods offer several advantages, including the ability to learn patterns or features in spectra and make predictions based on them, as well as improved performance and efficiency compared to other methods \citep{Casey2016}.
One advantage of the forward modeling approach is its interpretability, allowing for the examination of residuals and the identification of new systematics and explanatory variables. 
It also has the capability to handle missing data \citep{2023arXiv230202611A}. 
However, both generative and discriminative models have their limitations, including the risk of learning physically implausible relationships \citep{Hogg2019}. 
Additionally, they may not offer novel insights beyond our current understanding of the underlying physics.

In contrast, the direct discriminative approach can identify features in spectra that are correlated with stellar parameters. 
However, it may not capture all the relevant parameters and can be susceptible to systematic errors. 
These supervised-learning models are generally easier to train and better at avoiding overfitting, where the model becomes overly complex and fails to generalize to new data. 
Forward modeling approaches can address overfitting by incorporating a regularization term \citep{Casey2016}.
\cite{OBriain2021} introduced a hybrid generative domain-adaptation method that utilizes unsupervised learning on large spectroscopic surveys to transform simulated stellar spectra into realistic spectra. This method successfully calibrates synthetic data to match observations \citep{Wang2023b}, bridging the gap between theoretical models and practical observations. 
It also enables the identification of missing spectral lines in synthetic modeling \citep{OBriain2021}. 
This innovative approach has the potential to enhance data analysis techniques in stellar spectroscopy and other fields that rely on large datasets. 
Notably, the methodology employed in this study offers a balanced approach to stellar parameterization, neither relying solely on forward models nor discriminative models.
% \jd{cite yuan-sen transformer}

In this paper, we present a novel data-driven method called \xp, which enables the simultaneous estimation of stellar labels (\teff, \logg, \moh, and \aom) for red giant branch (RGB) stars using \gaia\ XP spectra combined with \apogee\ labels. 
Our approach lies between the forward modeling and direct supervised methods, leveraging the benefits of both.
The architecture of our model functions as a mapping from XP spectra to \apogee\ labels, but we incorporate \apogee\ spectra during training to enhance the model's performance. 
This combination allows us to exploit the rich information in both datasets effectively.
We demonstrate that \gaia\ XP spectra can accurately predict \teff\ and \logg\ due to their clear reflection in the overall spectrum profile. However, the estimation of metal abundance (\moh) and $\alpha$ abundance (\aom) from XP spectra is more challenging. 
Despite this challenge, our model can still achieve comparable precision in determining \moh\ and \aom\ as with LAMOST spectra, which typically have higher resolution of $\mathcal{R} \approx 1800$.
For \moh, the expected median absolute error (MAE) was estimated to be 0.1-0.2 dex for \gaia\ XP spectra \citep{Liu2012}. 
Although \cite{Andrae2022} reported a slightly higher MAE of 0.21 dex for \moh\ derived from \gaia\ XP spectra compared to \apogee, this information is still valuable, albeit at a qualitative level.

Recently, \cite{Rix2022} and \cite{2023arXiv230202611A} (henceforth \citetalias{2023arXiv230202611A}) developed an extreme gradient boosting ensemble model trained on \apogee\ \moh\ values and achieved a significantly improved median absolute error (MAE) of 0.06 dex when using only XP information. However, the estimation of $\alpha$-abundance (\aom) from XP spectra has been found to be a remapping between \aom\ and other parameters rather than a direct causal effect of $\alpha$ information \citep{Gavel2021}.

In our study, we demonstrate that our approach successfully derives meaningful \aom\ values for a large sample of approximately 23 million stars. 
This sample size is more than an order of magnitude larger than the current largest sample of \aom\, measurements, which consists of around 2 million stars observed with a resolution of $\mathcal{R}\sim 1800$ in LAMOST DR8. 
This substantial increase in sample size provides unprecedented statistical power for studying the Galactic enrichment history of the Milky Way.
An additional advantage of our data product, presented in this paper, is its independence from complicated selection effects introduced by cross-matching with multiple catalogs. 
Our published catalog is based solely on the selection function of \gaia, ensuring a homogeneous and all-sky dataset for studying the Galactic enrichment history of the Milky Way.

The subsequent sections of the paper are organized as follows:
Section \ref{sec:data} focuses on the dataset utilized for training \xp\ and provides insights into its composition and characteristics.
In Section \ref{sec:model}, we provide a comprehensive explanation of the \xp\ method, including a detailed description of the model architecture and the loss function.
In Section \ref{sec:result}, we evaluate the performance of \xp\ and present a catalog containing the labels and uncertainties of approximately 23 million red-giant stars obtained using \xp.
Finally, in Section \ref{sec:discussion}, we conclude the paper by discussing the implications of our results and addressing potential limitations and considerations associated with the use of our data product.
The resulting catalogs generated in this study have been published online \citep{jiadong_li_2023_8002699} and can be accessed at the following link: \url{https://zenodo.org/record/8002699}.
 
 %%%%%%%%%%%%%%%%%%%%%%%%%%%%%%%%%%%%%%%%%%%%%%%%%%%%%%%%%
%
%
%
%    Section: Data          
%
%
%
%%%%%%%%%%%%%%%%%%%%%%%%%%%%%%%%%%%%%%%%%%%%%%%%%%%%%%%%%
\section{Data} \label{sec:data}

In this Section we briefly introduce the Gaia \emph{XP} data  and their preprocessing, and the APOGEE training set.

\subsection{XP: the Gaia BP/RP low-resolution spectra}

The third data release of the \gaia\ mission (DR3) \citep{GaiaCollaboration2022} offers low-resolution aperture prism spectra \citep{DeAngeli2022, Montegriffo2022} for approximately 220 million stars. These spectra are obtained using the blue photometer (BP, 330-680 nm) and red photometer (RP, 640-1050 nm) instruments. The \gaia\ observation coverage includes a staggering 78 billion transits, and the processing pipeline of BP/RP spectra generates calibrations for each individual transit spectrum out of a total of 65 billion epoch spectra \citep{DeAngeli2022}. The final data product consists of more than two billion sources obtained by averaging the epoch spectra.

It's important to note that the XP spectra differ from classical spectra in terms of their representation. 
Instead of providing flux values corresponding to specific sampled wavelengths, the XP spectra are represented as a continuous function using a set of basis functions \citep{DeAngeli2022}. 
The continuous spectra are then encoded as an array of coefficients, with the first coefficients capturing the majority of the flux and the higher-order coefficients storing narrow spectral features \citep{DeAngeli2022}. 
This representation maximizes the information content of the XP spectroscopy data by efficiently representing the spectra with a reduced number of coefficients \citep{DeAngeli2022}.

\begin{figure*}[hbt!]
\centering
\includegraphics[width=\linewidth]{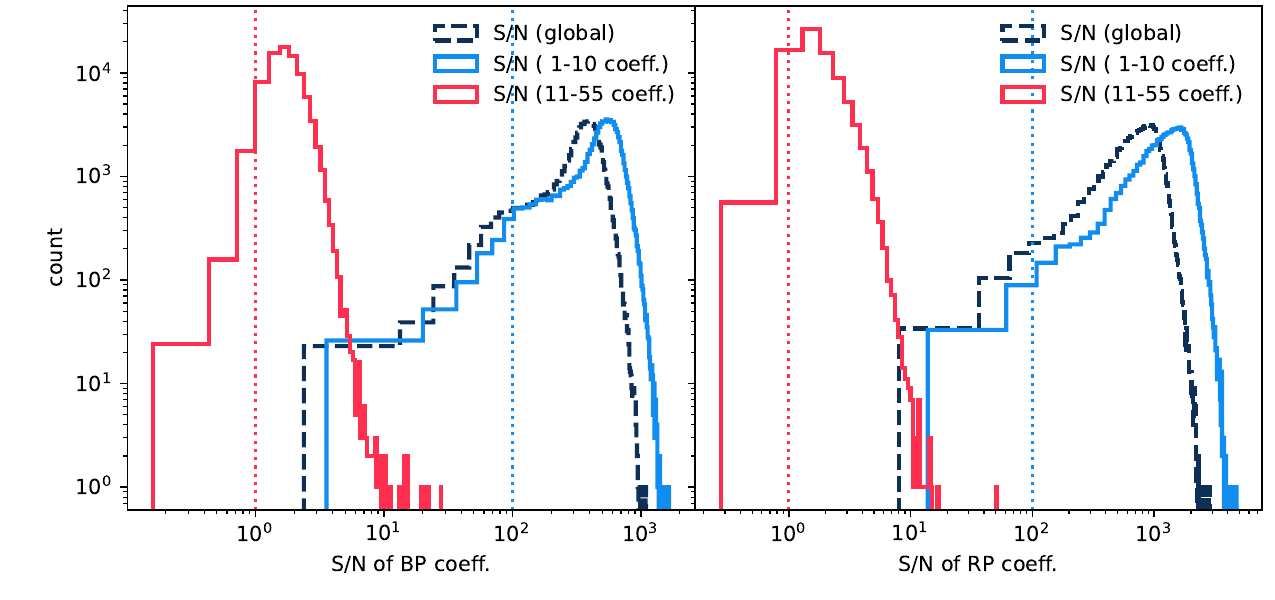}
\caption{Distribution of the signal-to-noise (S/N) \gaia\ XP coefficients for the sample, shown for the BP portion of the spectrum on the left, for the RP poertion of the spectrum on the right.  
The black dashed histograms represent the sample's distribution when the S/N is averaged across all 55 coefficients in BP and RP, respectively.  
The blue and red histogram show the analogous distributions considering only the first 10 coefficients (blue) and the remaining higher-order coefficients (red), respectively.  
The vertical lines of different colors in the figure represent the S/N thresholds employed in our training sample. 
Specifically, we consider spectra with S/N values greater than 100 for both the BP and RP coefficients (1-10), while for coefficients 11-55, a S/N threshold of 1 is applied.
% \HWR{..that we use in?}. 
}
\label{fig:snr}
\end{figure*}

\begin{figure*}[hbt!]
\centering
\includegraphics[width=\linewidth]{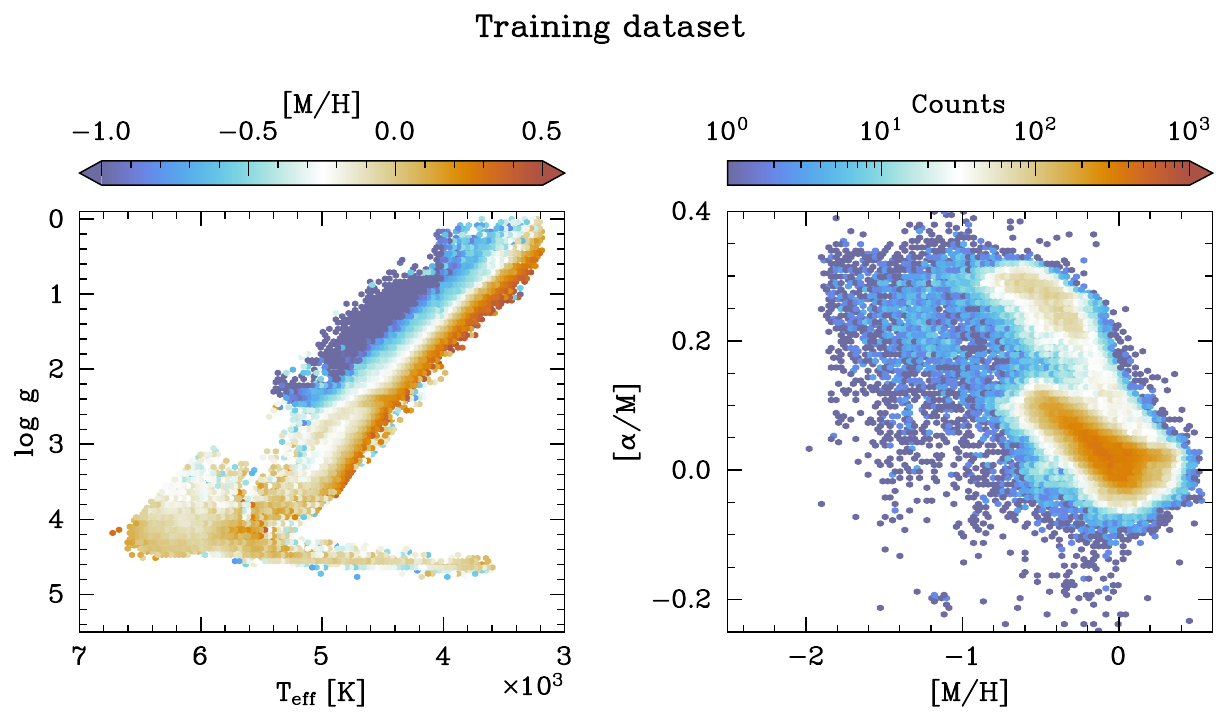}
\caption{\xp\ training set for this application, taken from SDSS  \apogee\ DR17. The left panel shows the \teff~--~\logg\ Kiel diagram of the training data, color-coded by \moh .
 The right panel shows the number density distribution of the \moh - \aom abundance diagnostics in the training set.
 \xp\ is designed to estimate \teff, \logg, \moh, and \aom\ values from XP spectra that are consistent with the SDSS \apogee\ training data.
}
\label{fig:train}
\end{figure*}

\subsection{Training datatset}\label{subsec:trdata}

For training, we use stars in common between the Gaia XP data and \apogee\ targets in the SDSS Data Release 17 \citep{Abdurrouf2022} and \gaia\ DR3 of interest, after some data cleaning.
First, we rule out stars with problematic flags ({\tt\string ASPCAPFLAG}) from the \textsl{ASPCAP}. We remove stars with  {\tt\string WARNING} or {\tt\string BAD} flags in the data model defined in \textsl{ASPCAP} for \teff, \logg, \moh, and \aom, as well as stars with flags with {\tt\string CHI2} and {\tt\string NO\_GRID}. 
The reason we did not use a strict flag cut (i.e., {\tt\string ASPCAPFLAG} $\neq 0$) to obtain a clean RGB training sample is that labels on parameter boundaries tend to be harder to estimate, and expanding the boundaries to main-sequence stars would move the labels of the giant stars of interest away from the boundaries.
Second, samples are selected using the following criteria as displayed in Table~\ref{tab:crit} to obtain both reliable stellar labels from \apogee\ and to eliminate stars without meaningful astrophysical parameters:
% \kw{Can we put the criterion in some other form like a table? This look odds to me now}
% \begin{enumerate}
% \item $3000 \leq$ T$_{\rm eff}$ $\leq 7000$~K; 
% \item $0 \leq \log ~g \leq 5.5$~dex;
% \item $-2.5 \leq {\rm [M/H]}  \leq 0.6$~dex; 
% \item $-0.1 \leq {\rm [\alpha/M]}  \leq 0.6$~dex
% \item
% $\log g > 4.2 - 1.5 \log(T_{\rm eff}/5777)$ and $\log g < 4.8$ dex  $\lor$
% \\
% $\log g < (5/3000) (T_{\rm eff}-3000)+0.5$ and $\log g < (5/3000) (T_{\rm eff}-3000)-1.9$;
% \item \eteff $\leq 50$~K; 
% \item \elogg $\leq 0.05$~dex;
% \item \emoh $\leq 0.02$~dex;
% \item \eaom $\leq 0.02$~dex;
% \item S/N of APOGEE $>200$.
% \end{enumerate}

\begin{table*}[hbt]\label{tab:crit}
% \centering
\caption{Criteria to deem stellar labels from SDSS/\apogee\ reliable, and eliminate stars without meaningful astrophysical parameters.}
\begin{tabular}{ll}
\centering
\textbf{Criterion} & \textbf{Range/Condition} \\
\hline
1. & $3000 \leq \text{T}_{\text{eff}} \leq 7000$ K \\
2. & $0 \leq \log g \leq 5.5$ dex \\
3. & $-2.5 \leq \text{[M/H]} \leq 0.6$ dex \\
4. & $-0.1 \leq \text{[$\alpha$/M]} \leq 0.6$ dex \\
5. & $\log g > 4.2 - 1.5 \log(\text{T}_{\text{eff}}/5777)$ and $\log g < 4.8$ dex $\lor$ \\
   & $\log g < (5/3000) (\text{T}_{\text{eff}}-3000)+0.5$ and $\log g < (5/3000) (\text{T}_{\text{eff}}-3000)-1.9$ \\
6. & $\sigma_{\text{T}_{\text{eff}}} \leq 50$ K \\
7. & $\sigma_{\log g} \leq 0.05$ dex \\
8. & $\sigma_{\text{[M/H]}} \leq 0.02$ dex \\
9. & $\sigma_{\text{[$\alpha$/M]}} \leq 0.02$ dex \\
10. & S/N of APOGEE $>200$ \\
\hline
\end{tabular}
\end{table*}

The condition \emph{5.} identifies stars that are eother on the main-sequence or on the RGB. 
% The condition 5 and condition 6 are logical relations of or, otherwise they are relations of intersection.

Third, we further apply the following conditions to select 142,130 stars with good \apogee\ labels, and we display the S/N distributions of coefficients in Figure \ref{fig:snr}. Here, the S/N is defined as the L2 norm of the given coefficients divided by the L2 norm of the corresponding errors. We find:
\begin{enumerate}
\item Most of the stars have high global S/N ($>100$) values for the coefficients.
\item Although the S/N of high-order coefficients (11-55), which contains abundance information, is lower than the first-order coefficients (1-10), most of them are larger than 1, indicating that there is still valuable information present.
\end{enumerate}

For the balance of \moh\ and \aom, we uniformly weight \aom\ and \moh\ in two dimensions. We re-sample stars into various \moh-\aom\ bins with specific bin edges. The bin edges for \moh\ are as follows: 2.0, -1.8, -1.6, -1.4, -1.2, -1.0, -0.9, -0.8, -0.7, -0.6, -0.5, -0.4, -0.3, -0.2, -0.1, 0, 0.1, 0.2, 0.3, 0.4, and 0.6. 
The bin edges for \aom\ are -0.2, -0.1, 0.0, 0.1, 0.2, 0.3, and 0.4.
In each bin, we perform random sampling of 1,000 stars with replacement from the selected sample, resulting in a total of 111,000 training data points.
This re-sampling process ensures a flat distribution of stars in the \moh \textit{vs} \aom\ space, allowing for better representation and balance in the training set.

% For intertal validation During the training process, we randomly choose 50\% of the above selected sample as the training data, and the remaining 50\% stars as the validation dataset. 
\subsection{Preprocessing}\label{subsec:preprocess}

Then we pre-processing the data before we train the \xp. 
We first concatenate the BP and RP coefficients to a 110-element array.
Then we use the value of $10^{0.5(15-G)}$ ($G$ is the \gaia\,$G$-band magnitude) as the normalization value, and divided by the 110-element XP array.

In cases where the \gaia\ XP coefficients exhibit a wide range of values spanning four orders of magnitude from first to high coefficients, it is found that the higher-order coefficients tend to cluster around similar values. 
In such scenarios, using the median and interquartile range as normalization measures can yield more robust results. 
Some works of data-driven methods to pre-process the spectra use standard normalization, performed by removing the mean and scaling to a unit variance (e.g., \citealt{zhang2020a}). 
However, outliers can often negatively affect the sample statistics, especially for noisy data. 
Hence we adopt the median value of each $j$th coefficient, $\mu_j$ as the centering value, and the coefficient range (the difference between the 25th quantile and 75th quantile) as the scale. 
Let $x_{i,j}$ be the $j$th coefficient of the $i$th spectra, then we define the scaling coefficient $\hat{x}_{i,j}$ for $j$th coefficient as 
\begin{equation}
   \hat{x}_{i,j} = \frac{x_{i,j} - x_{50}}{x_{75}-x_{25}}.
\end{equation}

Stellar labels are also scaled in the same way. Because the label measurements are more robust than high-order XP coefficients, here we choose the label range of scale as $y_{97.5}-y_{2.5}$, where $y_{m}$ is the $m$-th percentile value of label $y$.

%%%%%%%%%%%%%%%%%%%%%%%%%%%%%%%%%%%%%%%%%%%%%%%%%%%%%%%%%
%
%
%
%    Section: Method        
%
%
%
%%%%%%%%%%%%%%%%%%%%%%%%%%%%%%%%%%%%%%%%%%%%%%%%%%%%%%%%%

\section{Method: building the \xp\ model} \label{sec:model}

\begin{figure*}[htbp]
    \includegraphics[width=\linewidth]{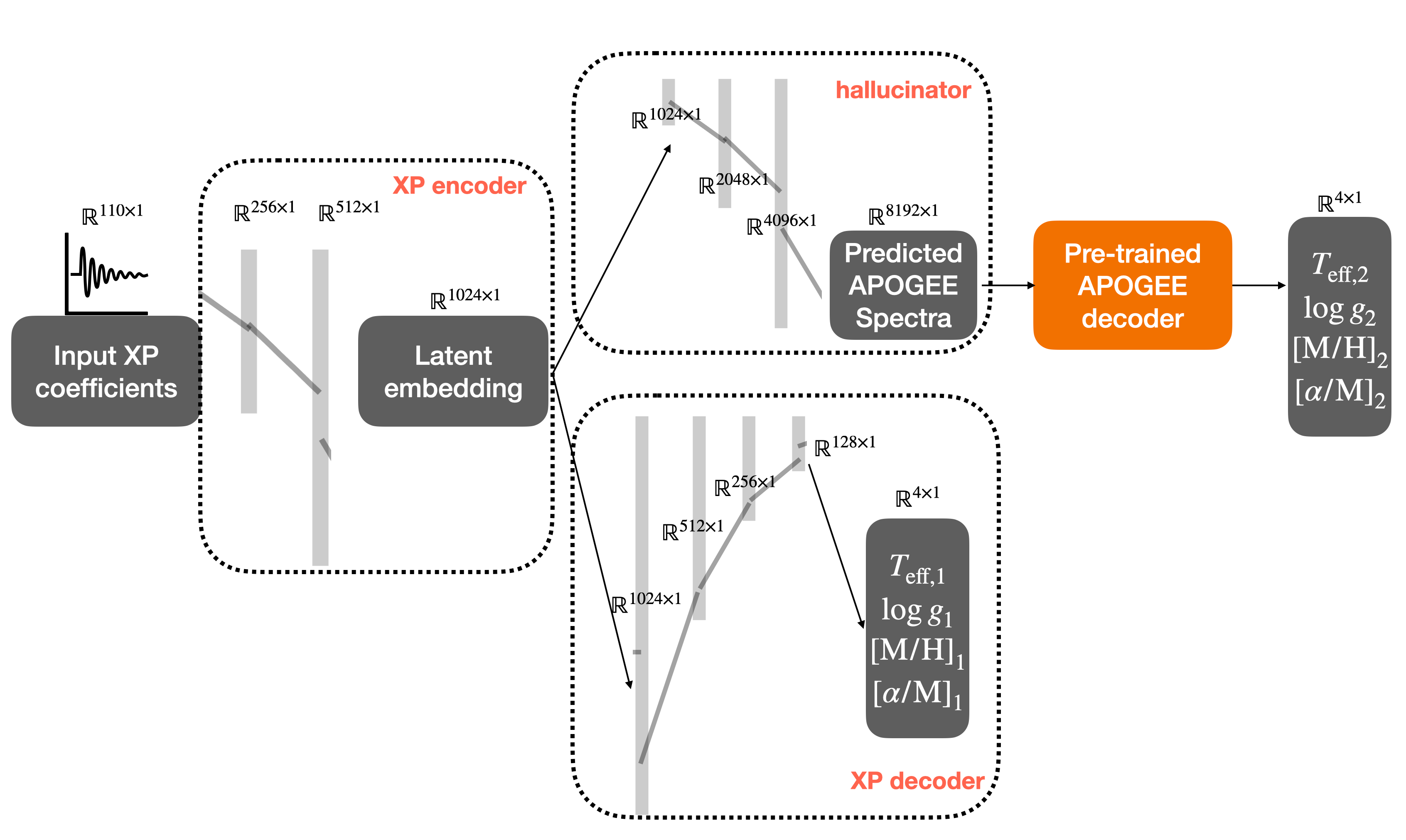}
    \caption{The architecture of the \xp\ model, consisting of four blocks: a pre-trained APOGEE decoder that takes APOGEE(-like) spectra to generate predictions for stellar labels; an XP encoder that generates 1024 shared latent variables from the $2\times 55$ XP coefficients; the \emph{hallucinator}, which generates APOGEE-like spectra from these embedded variables to be fed into the pre-trained APOGEE decoder; and the XP decoder that generates stellar label predictions from the XP encoder's 1024 latent variables. 
    During the training process, both sets of stellar label predictions, obtained from the XP decoder and the \emph{hallucinator}/APOGEE decoder, are utilized in the objective function of \xp.
    The fundamental concept behind the design of \xp\ is the prospect that incorporating the stellar label generation through the \emph{hallucinator} route will lead to a more robust and physically meaningful XP encoder following the training phase. 
    In the testing phase and for inference on other data, we only use the labels generated by the XP decoder. 
    This is because, once the model is well trained, the mean-square error between the \emph{hallucinator}'s predictions and the ground truth, as well as between the XP decoder's predictions and the ground truth, becomes approximately the same. 
    % In training, \xp\ both sets of stellar labels predictions, from the XP decoder \emph{and} from the hallucinator/APOGEE-decoder, will enter the objective function \HWR{True?}. 
    % The basic idea underlying the \xp\ layout is that the stellar label generation via the  \emph{hallucinator} route will result in a more robust and physically meaningful XP encoder after the training step.
    % \HWR{Perhaps state already here whether in the test step you will use the labels via the XP decoder route, via the hallucinator route, or both.}
    } 
    %The XP encoder maps low-resolution XP spectra (here, XP coefficients) to a latent space, incorporating information from high-resolution APOGEE spectra to enhance the XP spectra. The hallucinator network generates APOGEE-like spectra from the latent space representation, which are then decoded by both the pre-trained APOGEE decoder and a decoder specific to XP spectra.}
    \label{fig:model}
\end{figure*}

Our goal is to create a data-driven model that estimates stellar labels from XP spectra. 
To achieve this, we have developed a neural-network-based model, \xp, which leverages the rich abundance information of high-resolution \apogee\ spectra. 

The \xp\ model is constructed in four blocks as shown in Figure \ref{fig:model}: a pre-trained APOGEE decoder, an XP encoder, a component dubbed the \emph{hallucinator}, and an XP decoder. 
% \kw{The main idea behind the architecture is to leverage information in APOGEE spectra during training. In the most basic scenario, we only use GAIA spectra to predict the stellar labels, which does not put any constraints on how the prediction is done. By requiring the network to also predict the APOGEE spectra during training, we force the network to be consistent with how stellar label is derived from APOGEE spectra. This imposed consistency check discourages the network to overfit to the GAIA data, hence improving the generalization performance.}
%\jd{
The key idea behind the \xp\, architecture is to fully exploit the rich abundance information contained in \apogee\ spectra to improve stellar parameter estimates from \gaia\ XP spectra. 
In a vanilla approach, we might only use \xp\ XP spectra as input to predict the stellar labels with an encoder and decoder. 
While this approach is valid, it does not impose any constraints on how the prediction is made, e.g. that the prediction draws on physically meaningful parts of the XP spectra. Training a neural network involves determining the weights and biases of a flexible model that fits the training data. 
Within the set of possible parameter combinations that fit the data well, some may result in overfitting, exhibiting unphysical behavior outside the training data. 
Our model requires the introduction of extra constraints that entail matching the APOGEE spectra, resulting in more robust results on unseen data\footnote{The authors are aware that a formal distinction between a method that ``predicts'' and one that determines or ``measures'' stellar labels is a subtle and far reaching issue, beyond the scope of the paper}. 
% This may increase the risk that the overfits the data, or merely ''predicts'' the labels, rather than estimate or determine them from the \gaia\ data
% \xp\ addresses this issue by requiring during training that the model also predicts the corresponding \apogee\ spectra from the XP spectra, in addition to predicting the stellar labels. 
This forces the model to learn a representation of the \xp\  spectra that is consistent with how stellar parameters are derived from \apogee\ spectra.
%}

%\jd{
By incorporating the hallucinator component, we effectively modify the
XP encoder component (that is common to both subsequent branches), as it must be able to allow the prediction of \apogee -like spectra, where stellar label are determined from physically well-defined spectral features. We anticipate that this discourages the model from overfitting when mapping XP spectra to stellar labels alone, thereby improving the test-step performance, when applied to new XP spectra.
By leveraging both XP and \apogee\ spectra simultaneously during training, \xp\ exploits explicitly the rich information in \apogee\ spectra. As we show below, this indeed substantially improves the stellar label estimates, in particular the \aom\ abundance measurements, from the low-resolution XP spectra.
%}

As Figure \ref{fig:model} illustrates, \xp~ starts with the XP encoder that maps the low-resolution XP spectra to a latent space. 
Then, the {\it hallucinator} network generates an \apogee -like spectrum from the this latent space representation. This spectrum then gets mapped to stellar labels, using a decoder that has been pre-trained on real \apogee~ spectra.
In a second branch of \xp\, , the initial latent embedding from the encoder gets mapped to the stellar labels directly via an XP decoder.

\begin{figure*}[hbt!]
\centering
\includegraphics[width=0.99\linewidth]{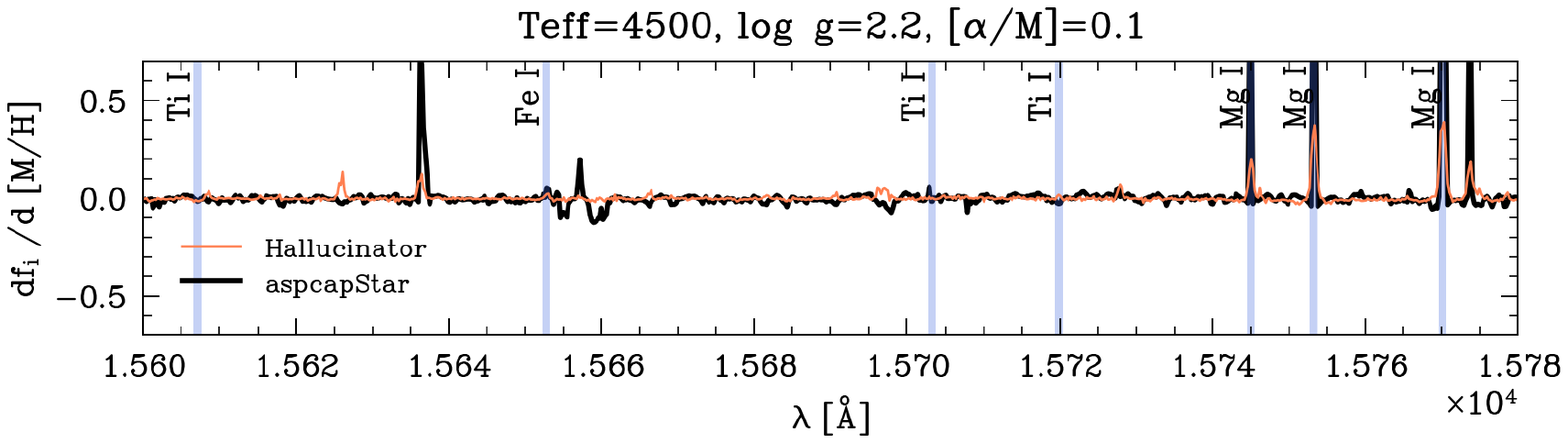}
\caption{Example of a predicted APOGEE-like gradient spectrum produced inside the hallucinator, compared to an actual APOGEE gradient spectrum.
% Note that this prediction is never confronted with actual APOGEE spectra. 
% It just has to be a spectrum from which the APOGEE pipeline produces the correct stellar labels.
During training, the hallucinator predictions are not directly compared to real APOGEE spectra. 
Rather, the hallucinated spectra only need to produce the correct stellar labels when fed into the pre-trained APOGEE pipeline decoder. 
This allows the hallucinator to learn a mapping that captures relevant spectral features correlated with the stellar labels.}
\label{fig:hallucinated_spectra}
\end{figure*}

Note that we do not require explicitly that the spectral features in the hallucinated APOGEE-like vary with changes in label exactly as expected ``from physics".  
Instead, the hallucinated spectra must only generate accuarate stellar labels when inputted into the pre-trained APOGEE pipeline decoder, originally trained on genuine \apogee~ spectra. 
%This allows the hallucinator to learn a mapping that captures relevant spectral features correlated with the stellar labels, which are shown as the strong gradient features as shown in Fig.~\ref{fig:hallucinated_spectra}.
Nonetheless, the hallucinated gradient spectra (i.e. the spectra's derivatives with respect to some stellar label), show that the hallucinated \apogee\ spectra vary, say with [M/H], at the wavelengths where physics-based spectral models expect them to. 
This is shown in Fig.~\ref{fig:hallucinated_spectra}, and demonstrates the hallucinator's ability to generate realistic synthetic spectra containing meaningful spectral information related to the stellar parameters.

In each block, we chose to use Multilayer Perceptron (MLP) as our feature extractor, rather than other powerful feature extraction techniques like convolutional neural networks (CNNs). 
We did this mainly because MLP is both versatile and simple. 
MLP is capable of learning non-linear relationships between input features, which makes it well-suited for problems where the relationships between the features are complex and difficult to capture using linear models. While other feature extraction techniques, such as CNNs, can be very powerful, they may not always be necessary for every problem. 
In our case, we found that MLP provided a good balance of performance and simplicity, allowing us to achieve good results without adding unnecessary complexity to our model. 

While we may have qualitative reason to expect that the {\it hallucinator} enforces a better latent embedding in the XP encoder in training, any improvements in the prediction of stellar test labels can only be estimated empirically, as we do below.
%Our approach allows us to extract high-level features that are relevant to the task of deriving stellar labels from the XP spectra.  By combining this information with the low-resolution XP spectra, we can improve the accuracy of our model. Additionally, we can use existing knowledge and expertise gained from decoding APOGEE spectra, rather than having to train a new decoder from scratch for the XP spectra.

\subsection{Objective function}

During the training process, we optimize an objective function and determine the model parameters $\mathbf{\theta}$. We denote the  $i^{\mathrm th}$ target's \apogee~ labels as $\hat{y_i}$, and their uncertainties as $\sigma_i$; and we denote the predicted labels from the XP decoder as $y_{{\rm model1},i}$, and those from the hallucinator and pre-trained \apogee\ decoder as $y_{{\rm model2},i}$ . 
The overall data-model variance term $s_i$ described as 
\begin{equation}
    s_{i}^2 = \max{(\epsilon, \sigma{_i}^2 + s_{{\rm model1},i}^2)}.
\end{equation}
%where $\sigma_i$ is the uncertainty of each \apogee\ training lable, and  $s_{{\rm model1},i}$ is the corresponding uncertainty of the model prediction $y_{{\rm model1},i}$. 
For stability in the training, we introduce a floor 
$\epsilon$ ($10^{-6}$) to prevent values of excessive $1/s_{i}$.

With these definitions, we can spell out the loss function, which penalizes deviations of predicted labels from true labels, and incorporates regularization terms to encourage model parameter sparsity:

% \begin{equation}\label{eq:loss function}
% \begin{aligned}
% L(\hat{y}, y_{\mathrm{model1}}, y_{\mathrm{model2}}) &= \frac{1}{N} \left[ \sum_{i=0}^{N} \frac{\alpha_1}{2} \frac{[\hat{y} - y_{\mathrm{model1},i}]^2}{s_{i}^2} + \frac{\alpha_2}{2} \frac{[\hat{y} - y_{\mathrm{model2},i}]^2}{s_{i}^2} \right. \\
% & \quad \left. + \frac{1}{2} \sum_{i=0}^{N} \ln s_{i}^2 + \Lambda_1 \|\pmb{\theta}\|_{\mathrm{L1}} + \Lambda_2 \sum_{l=1}^{4} D_{\mathrm{KL}}(\chi_l \| \mathcal{N}(0,1)) \right].
% \end{aligned}
% \end{equation}
%
\begin{equation}\label{eq:loss function}
\begin{aligned}
L(\hat{y}, y_{\mathrm{model1}}, y_{\mathrm{model2}}) &= \frac{1}{N} \biggl[ \sum_{i=1}^{N} \sum_{k=1}^{2} \frac{\alpha_k}{2} \frac{[\hat{y} - y_{\mathrm{model}k,i}]^2}{s_{i}^2} \\
&\quad + \frac{1}{2} \sum_{i=1}^{N} \ln s_{i}^2 + \Lambda_1 \|\pmb{\theta}\|_{\mathrm{L1}} \\
&\quad + \Lambda_2 \sum_{l=1}^{4} D_{\mathrm{KL}}(\chi_l \| \mathcal{N}(0,1)) \biggr].
\end{aligned}
\end{equation}

Here, $N$ represents the size of the training data. $\alpha_1$ and $\alpha_2$ denotes the relative weights for the loss contributed by the XP decoder and the APOGEE decoder. A good representation requires the decoder to predict the same stellar label both directly from the XP spectra and first from the XP spectra to APOGEE spectra, then decode to stellar label, hence we have both terms in our loss. 
The loss function comprises a weighted total of the losses computed by both decoders, the weights of which are determined by the relative importance of each decoder for the task at hand. 
% \HWR{Can you just say a bit more about the weighing? What did you do in practice?}
In actuality, these weights are obtained through a hyperparameter grid search, resulting in assigned values of 0.4 for the \emph{Hallucinator} and 0.6 for the XP decoder.
% They are determined to be $\alpha_1 = 0.6$ and $\alpha_2 = 0.4$through a grid search.
% \HWR{Can you say more about the weights? Or say, that you will say more later..}  
The term with $\ln s_{i}$ is to ensure the model does not just arbitrarily increase the variance of the model to reduce the loss.
$\Lambda_1$ is a regularization parameter, and $\|\pmb{\theta}\|_{\rm L1}$ is the L1-norm of sum of absolute values of the components of model parameter $\mathbf{\theta}$, and regularization term $ \Lambda_1 \|\pmb{\theta}\|_{\rm L1}$ encourages parameters to take on zero values. 
% In eq.\ref{eq:loss function}, we define the variance term $s_i$ as:
% %
% \begin{equation}
%     s_{i}^2 = \max{(\epsilon, \sigma{_i}^2 + s_{{\rm model1},i}^2)},
% \end{equation}
% %
% where $\sigma_i$ is the error associated with the $i$th target label from \apogee, and $s_{{\rm model1},i}$ is the corresponding uncertainties associated with the predicted labels $y_{{\rm model1},i}$. 
% And $\epsilon$ is a small value of $10^{-6}$ used to clamp the variance value $s_i$ for stability during training. 

To further encourage the model to estimate accurate uncertainties, we add a Kullback-Leibler (KL) divergence penalty to the loss function. Specifically, we measure the divergence between the distribution of the uncertainty-normalized difference $\chi\equiv{(\hat{y}-y_{{\rm model1}})}/{\sqrt{\sigma^2 + s_{{\rm model}}^2}}$, and the standard normal distribution $\mathcal{N}(0,1)$.
Ideally, the probability density distribution of $\chi$ should be distributed as $\mathcal{N}(0,1)$, if the predicted $s_{{\rm model}}$ are accurate. We assume the uncertainty-normalized difference $\chi$ follows a Gaussian distribution $\mathcal{N}(\mu_{\rm res}, \sigma_{\rm res})$, with $\mu_{\rm res}$ and $\sigma_{\rm res}$ as the mean and standard deviation of the Gaussian distribution. In the case at hand, the KL divergence is then given by:
\begin{equation}
D_{\rm KL} (\chi | \mathcal{N}(0,1)) = \ln{\frac{1}{\sigma_{\rm res}}} + \frac{\sigma_{\rm res}^2 + \mu_{\rm res}^2}{2} - \frac{1}{2},
\end{equation}

\begin{figure*}[hbt]
    \centering
    \includegraphics[width=\linewidth]{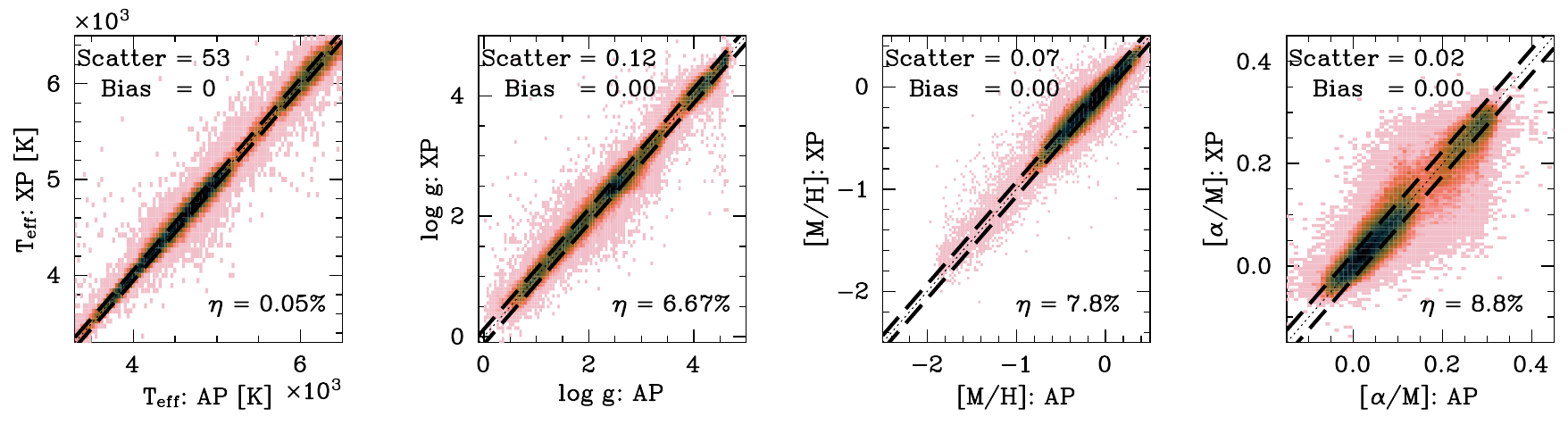}
    \caption{Validation of \xp-derived labels from \gaia\ XP spectra (XP) \emph{vs.} \apogee~labels (AP). 
    From left to right, the panels show the results of two-fold cross-validation drawn from the training set for  \teff, \logg, \moh, and \aom, respectively. 
    The (small) scatter and (negligible) bias are indicated in each panel, and $\eta$ is defined as the outlier rate given be Eq.~\ref{eq:eta}.
    The pairs of the dashed lines indicate the one-sigma deviation in the difference between the compared labels. 
    The color denotes the logarithm of the number density.
    }
    % \HWR{What are the dashed lines? What is $\eta$? }
    \label{fig:compare4d}
\end{figure*}

\begin{figure*}[hbt]
    \centering
     \includegraphics[width=\linewidth]{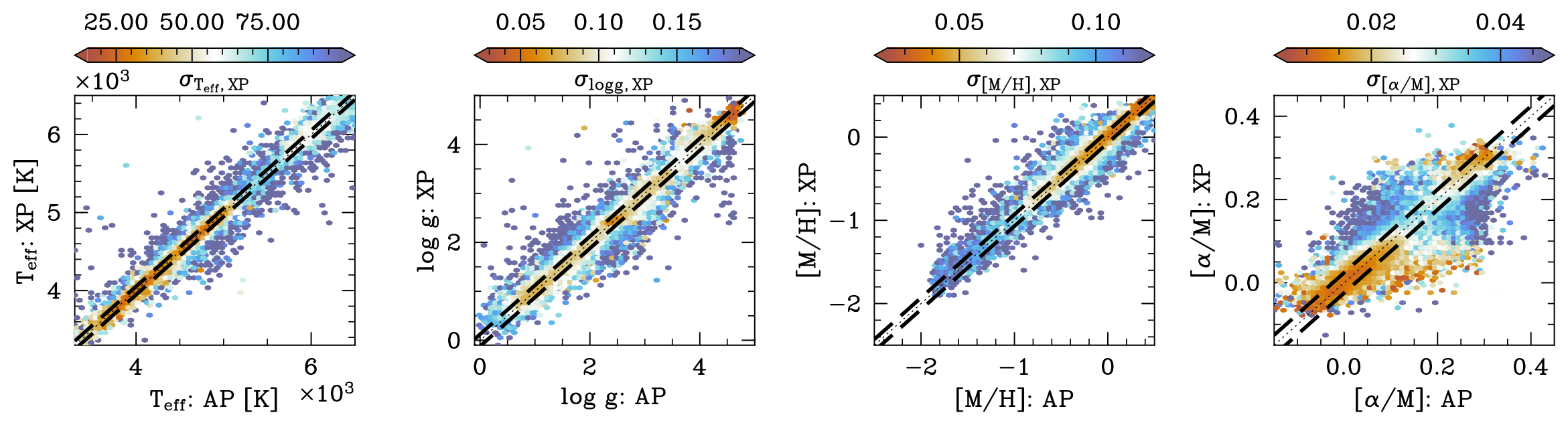}
    \caption{Validation of \xp-derived labels from \gaia\ XP spectra (XP) \emph{vs.} \apogee~labels (AP). This Figure is analogous to Fig.~\ref{fig:compare4d}, except that the color coding now denotes \xp 's estimate of the label prediction's uncertainty. Most objects with more discrepant \xp\ \emph{vs.} \apogee\  label estimates have correctly identified larger \xp\  label uncertainties.
    %The validaiton results are color-coded according to the corresponding errors of the labels derived from \xp. In each panel, the top plot is the validation of the labels derived from \xp, as a function of the actual label from \apogee;The bottom plot is the residuals, their residuals versus the true label of \apogee, and the white dashed line is the 1-sigma of the residuals. 
   }
    \label{fig:compare4d_err}
\end{figure*}

By adding the KL divergence term to the loss function, the model is penalized when the predicted distribution of uncertainty-normalized residuals deviates from the standard normal distribution. This encourages the model to produce more accurate uncertainties, as it must minimize the KL divergence in addition to minimizing the deviation between the predicted and true values. 
Overall, adding a KL divergence penalty to the loss function can help the \xp~ to estimate accurate uncertainties by ensuring that the predicted uncertainties are as close as possible to the true uncertainties, and that the predicted distribution of uncertainty-normalized residuals follows a standard normal distribution, and we will show the the error analysis in Section \ref{sec:result}.

\begin{figure*}
    \centering
     \includegraphics[width=\linewidth]{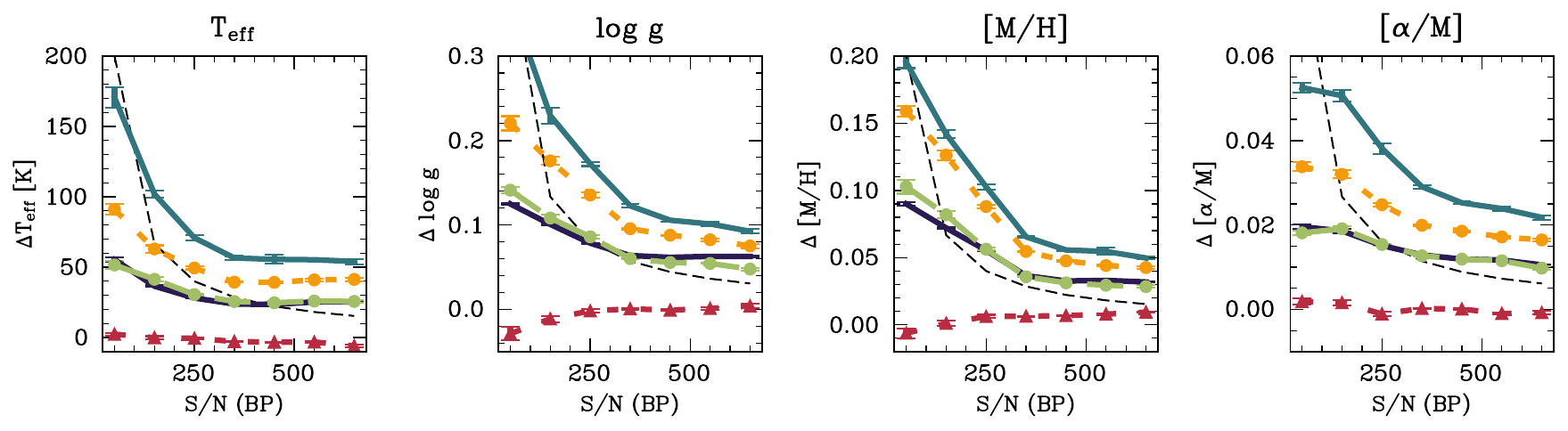}
    \includegraphics[width=\linewidth]{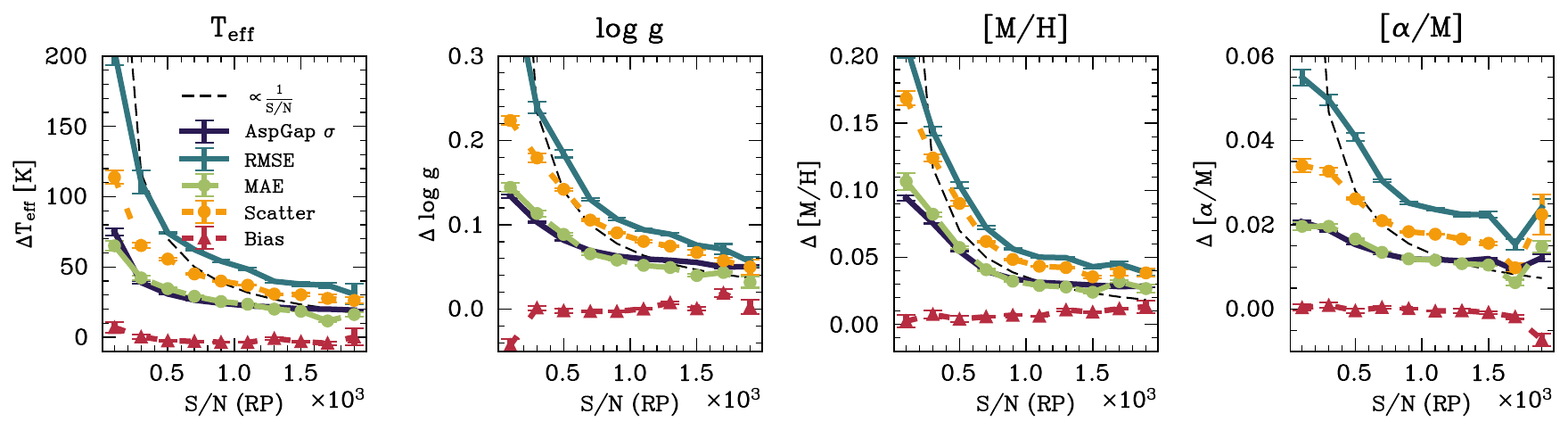}
    \caption{Quality of the labels generated by \xp , as a function of the input XP spectras' S/N, shown for \teff, \logg, \moh, and \aom from left to right. 
    In all panels, the solid black curves denote the formal errors generated by \xp. The teal, green, orange and red lines denote the cross-validation values for different mneasures of precision and accuracy: the root-mean-square deviation (RSME), the mean absolute deviation (MAE), the scatter (standard deviation), and the bias, respectively. The thin dashed line indicates the naive 1/(S/N) expectation. The top row shows this as a function of the BP S/N, the bottom row as a function of the RP S/N.  The cross-validation results approximately follow the S/N-scaling expectations for \teff, \logg, \moh , but less so for \aom .
    }
    \label{fig:error}
    \end{figure*}
    
\begin{figure*}
    \includegraphics[width=\linewidth]{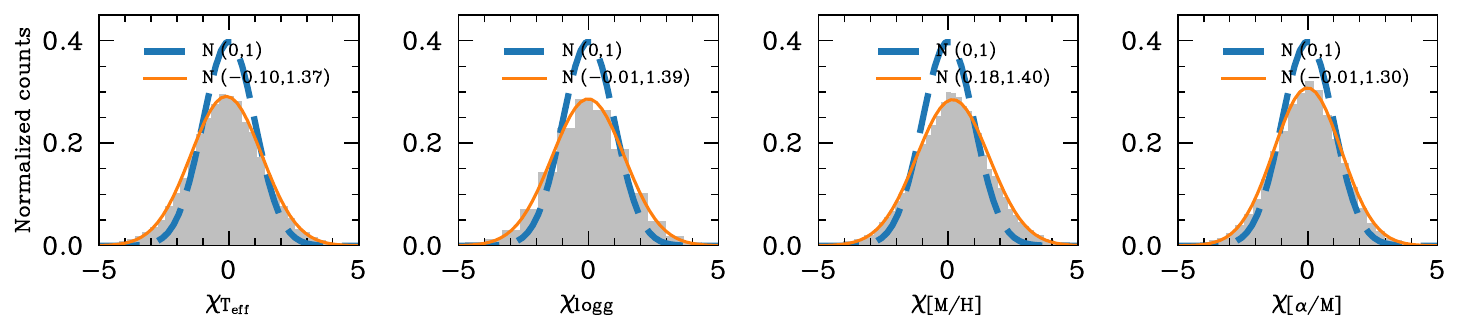}
    \caption{Fidelity of the label uncertainty estimates. The panels show the  probability density distribution of the uncertainty-normalized differences ($\chi$) between \xp~ and \apogee\ labels. 
    In each panel, the blue dashed curve represents the standard normal distribution $\mathcal{N}(0,1)$, and the orange solid line denotes the best-fitting normal distribution of $\chi$. The formal uncertainties are meaningful -- also for \aom~--  but about 30\% underestimated.
    % \HWR{perhaps random question, just to make sure: all these plots are for the FULL sample not for the giant sample shown in Figure 9?}
    % Yes, Figures 4, 5, 6, and 7 present the results of a 2-fold cross-validation using the training sample described in the Data section.
    By conducting these validations, we have access to the ground truth labels, which allows us to evaluate various performance metrics such as MAE ($\chi$), and others.
    }
    \label{fig:chi}
\end{figure*}

% \vskip 0.5truein
% \HWR{I think that there should be a mini-subsection or a paragraph on the "actual running of \xp ". Is it numerically trivial? Is it stable? Where does the code live?}

\subsection{Running \xp}

The implementation of \xp\ is numerically straightforward and stable. The core \xp\ model is built using {\tt\string PyTorch} and can be trained on a single NIVIDIA V100 GPU. 
Prediction of stellar labels on new XP spectra during inference is efficient, taking approximately 0.2 milliseconds per star on a single GPU.

We have open-sourced the full \xp\, code at \url{https://github.com/jiadonglee/aspgap} to allow others to replicate our results, extend the model for new applications, and deploy it for making predictions on large datasets. 
During training, we use the {\tt\string Adam} optimizer with an initial learning rate of 0.001. 
Training is run for 1,000 epochs with early stopping based on the validation loss. 
We do not find any significant numerical instabilities during training or inference. 
The model implementation and train/inference procedures are encapsulated in Python scripts, making it easy to apply \xp\, to new datasets from a programmatic interface or script.

%%%%%%%%%%%%%%%%%%%%%%%%%%%%%%%%%%%%%%%%%%%%%%%%%%%%%%%%%
%
%
%
%    Section: Results and validation        
%
%
%
%%%%%%%%%%%%%%%%%%%%%%%%%%%%%%%%%%%%%%%%%%%%%%%%%%%%%%%%%

\section{Results and validation}\label{sec:result}

In this Section we present the results of \xp\ the training and application just described.  After an initial internal validation, we present a stellar label catalog for a large sample of RGB stars, where we deem our results to be particularly robust and pertinent to astrophysical applications. We then present an extensive comparison with external data sets, followed by some ``astrophysical'' plausibility tests that imply that the determination of $[\alpha/M]$ (presented here for the first time for XP spectra) for giants is meaningful.

\begin{figure*}[htbp]
    \centering
    \includegraphics[width=\linewidth]{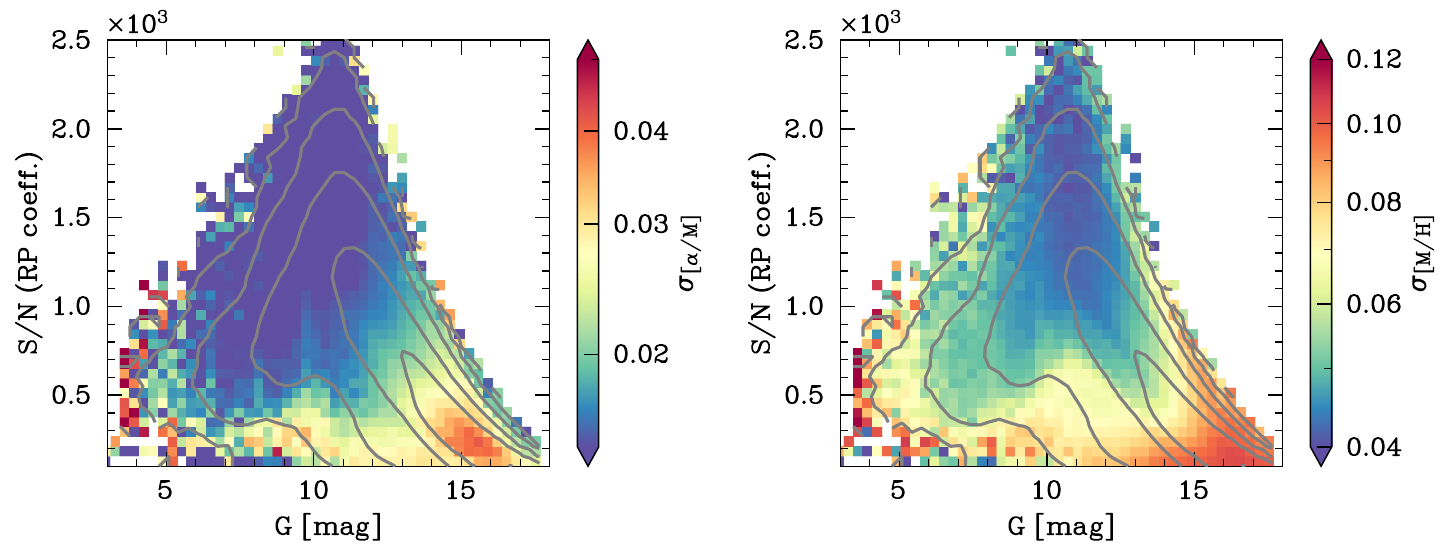}
    \caption{Dependence of the uncertainties from \xp~ for \moh~(right) and \aom~(left), shown as a function of the sources' G magnitude and S/N of the RP coefficients (see Fig.~\ref{fig:snr}. 
    The color indicates the mean variance in \aom~ and \moh, while the contours represent the number density of the sample, derived using a Gaussian kernel density estimate.
    % \HWR{Questions: 1) given that much of the abundance sensitivity is on the blue side, why did you choose the SN  in RP (not BP)? 2) What will the reader learn that is not part of the previosu figures?}
    % \jd{(Yes the abundance sensitivity of BP is indeed important, our analysis in Figure 6 demonstrates that the RMSE and scatter as a function of S/N in RP exhibit a closer relationship to 1/SN (as indicated by the dashed line), particularly for \teff, \logg, and \moh. 
    % Since these four labels are strongly correlated, we can infer that S/N in RP serves as a more effective indicator for diagnosing error estimations. The use of S/N in RP as an indicator provides a balance between these factors, it can be seen as a compromise.)
    }
    \label{fig:gmag_snr}
\end{figure*}

\subsection{Self-validation and error analysis}\label{subsec:error}
% Fig. \ref{fig:train} shows the \teff--\logg~ Kiel diagram of the training labels from the \apogee and the labels derived from \xp.
% We demonstrate that the cross-validated labels in the \teff--\logg diagram illustrate that \xp, can estimate \teff, \logg, and \moh as well as the \apogee training labels, as shown in the top row of Fig. \ref{fig:train}. Although, the biomodality of \aom\ is not as clear as the abundance of the \apogee.

For internal validation, we evaluate the performance of our \xp\, method by repeated two-fold cross-validation (CV) on the APOGEE training sample.
This involves dividing the training data into two equal subsets, and then iteratively swapping the subsets as training and validation sets.
Fig. \ref{fig:compare4d} displays these CV results for our four stellar labels ( \teff\textsubscript{XP}, \logg\textsubscript{XP}, \moh\textsubscript{XP}, and \aom\textsubscript{XP}) inferred by \xp~ from XP coefficients, plotted against the corresponding APOGEE labels, taken to be the ground truth. 
The \emph{rms} scatter  of the four labels are 48~K for \teff, 0.12~dex for \logg, 0.07~dex for \moh, and 0.03~dex for \aom, respectively.
The comparison immediately affirms that the performance of our stellar label prediction from the \gaia, low-resolution spectra is comparable to the quality of data-driven labels from e.g. LAMOST \citep{Ho2017}. 
For bright stars ($G<14$) the scatter is even slightly lower for all four labels (in particular for \aom\textsubscript{XP}): 46~K, 0.10~dex, 0.06~dex, and 0.021~dex, respectively.
For all labels, we define an outlier rate $\eta$ over the full sample of size $N$ via
\begin{equation}\label{eq:eta}
\eta = \frac{1}{N}~\times \left|~ \{\mathrm{deviation}_i : |\mathrm{deviation}_i| > 1 \}~ \right|
%{\sum_{i=1}^{N} \left|\frac{{\text{{res}}_i}}{{\text{{out\_bound}}}}\right| > 1}, 
\end{equation}
where $\text{{deviation}}_i$ is the difference between the ground truth and the predicted values for the $i^\mathrm{th}$ data point divided by a specified threshold for determining outliers, $\text{{out\_bound}}$. 
% \HWR{How do you set the threshold?}
The determination of the threshold for \teff\ and \logg\ is based on residuals exceeding 15\% of the ground truth values. 
Concerning \moh, outliers are identified when the residuals deviate by 0.15 dex from the ground truth. Similarly, for \aom, outliers are detected when the residuals deviate by 0.05 dex from the ground truth.

Remarkably, the outlier rate for the predicted \teff\ stands impressively low at 0.05\%, signifying that only a minute fraction of the predicted \teff\ values deviate significantly from the ground truth values.
On the other hand, the outlier rate for the predicted \logg\ is higher at 6.67\%, suggesting a relatively larger proportion of predicted \logg\ values that fall outside the specified threshold for determining outliers. 

We also show the comparison color-coded by \xp-derived uncertainties in Figure~\ref{fig:compare4d_err}.
We find that the \xp-inferred labels that differ more from the true \apogee have been assigned larger \xp~uncertainties, affirming that these error estimates are meaningful.

Fig.\ref{fig:error} further illustrates how the uncertainties change with signal-to-noise ratio (S/N), showing the formal \xp~ uncertainties, as well as RMSE, MAE, scatter, and bias at different S/N. 
As the S/N increases, the \xp~uncertainties as well as the RMSE, MAE and scatter decrease rapidly, as expected.
The scatters are $\sim$ 40~K, 0.10~dex, 0.05~dex, and 0.02~dex for \teff, \logg, \moh, and \aom~ for a (RP coefficient) S/N of $\sim 800$).
For the best S/N ($>$1500) in the RP coefficients, the scatters are 30~K, 0.06~dex, 0.04~dex, 0.01~dex for \teff, \logg, \moh and \aom, respectively. These scatters are comparable to those published for data-driven stellar labels derived from high-resolution \apogee~ spectrum \citep{Ness2015, Casey2016, Ting2019a}. 

The label uncertainties derived by \xp~ (formal error) as a function of S/N are also shown in Fig.\ref{fig:error}. 
The formal errors from \xp~decrease with S/N,  similar to MAE. 
Beyond S/N of $\gtrsim 500$ the uncertainties of all labels reach a floor, indicating that they are dominated by systematic errors rather than random uncertainties.

We further validated the uncertainty estimates by looking at the $\chi^2$ statistic.
Fig.\ref{fig:chi} presents the uncertainty-normalized difference ($\chi$) between the true labels and those predicted by the model. 
Compared to the standard normal distribution $\mathcal{N}(0,1)$, we find that the \xp\ uncertainty for \teff, \logg, and \moh is underestimated by about 30\% if we assume that the \apogee\ labels are exact. 
% For \aom, the \xp\--errors are more convincing, because the difference with the standard deviation of the standard normal distribution is only $\sim$ 0.1.
% We can draw the following conclusions from in Fig.\ref{fig:cat}. 
% First, although the errors derived from XP is an order of magnitude higher than that claimed by ASPCAP, the \moh\ gradient is almost the same as the \apogee\ training data set in Fig.\ref{fig:train}.
% Second, the uncertainties of \moh\ and \aom\ are higher for metal-poor stars (\moh$<-1$) than for metal-rich stars (\moh$>-1$).
% The \xp\ errors of abundances (\moh\ and \aom) of metal-poor stars are at least twice as high as that of metal-rich stars. 
% It should be mentioned that the median error of the \apogee\ labels used for training is also more than 2 times higher for the metal-poor errors than for the errors of metal-rich stars. 
% In the metal-poor regime, \moh\ and \aom\ are highly degenerate, as reported in [cite], thus the larger errors of the \xp-derived errors could be caused by the large uncertainties of the \moh\ estimation.
% For metal-rich stars, the errors of high \aom\ stars (\aom$>0.1$) are larger than those of low--$\alpha$ stars (\aom$<0.1$). 
The uncertainties of \xp\ predictions are influenced by both the signal-to-noise ratio (S/N) and the $G$-band magnitude of the stars, as shown in Figure \ref{fig:gmag_snr}. 
For stars with high S/N and bright magnitudes ($G < 12$), the uncertainties are generally low. However, it is important to note that label precision is best predicted by the spectral S/N rather than by the magnitude itself. 
This implies that the quality of the spectra, as indicated by the S/N, plays a crucial role in determining the accuracy of the \xp\ predictions. 
Therefore, when aiming to select samples with small uncertainties in their \xp\ predictions, it is essential to take into account not only the $G$-band magnitude of the stars, but focus on ensuring a sufficient S/N.

%%%%%%%%%%%%%%%%%%%%%%%%%%%%%%%%%%%%%%%%%%%%%%%%%%%%%%%%%
%
%
%
%    Section: Catalog        
%
%
%
%%%%%%%%%%%%%%%%%%%%%%%%%%%%%%%%%%%%%%%%%%%%%%%%%%%%%%%%%
\subsection{The Catalog of RGB Stars}

\begin{figure*}[htbp]
    \centering
    \includegraphics[width=0.6\linewidth]{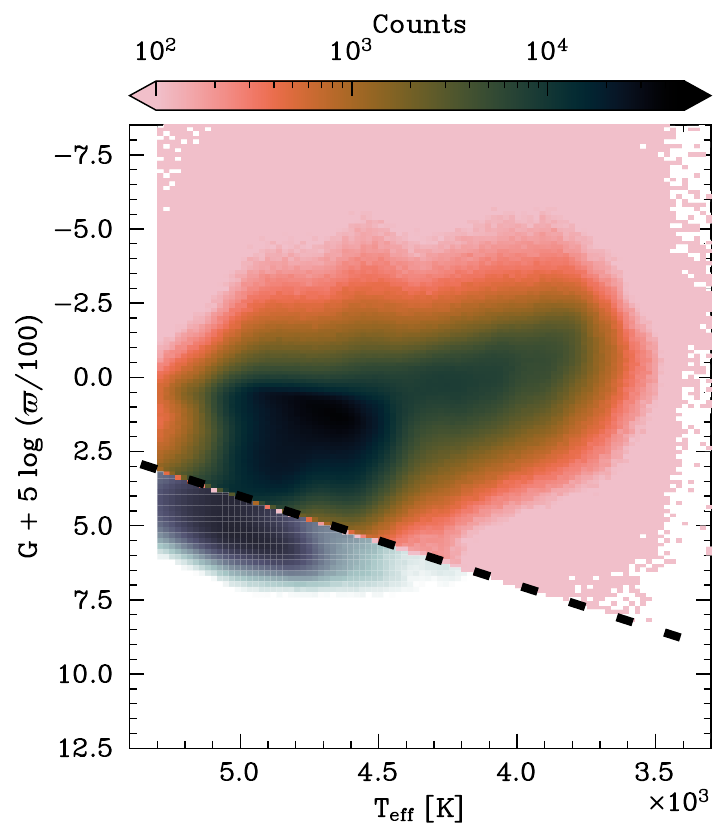}
    \caption{Definition of the \textit{RGB sample}, for which the \xp~label estimates are most precise and robust. 
    The sample cuts are  illustrated in the H-R diagram (color-coded by the number density of the \gaia\ XP sample) with two conditions $T_{\rm eff}<5300\,K$ and $\varpi \cdot 10^{G/5} < 10^{(-0.003 T_{\rm eff} + 19)/5+10}$, where pseud-luminosity (left-hand side) in the second cut was chosen to remain well defined even for $\varpi \le 0$. They lead to a sample of 23 million objects.}
    \label{fig:hrdcut}
\end{figure*}

\begin{figure*}[hbt!]
    \centering
    \includegraphics[width=\linewidth]{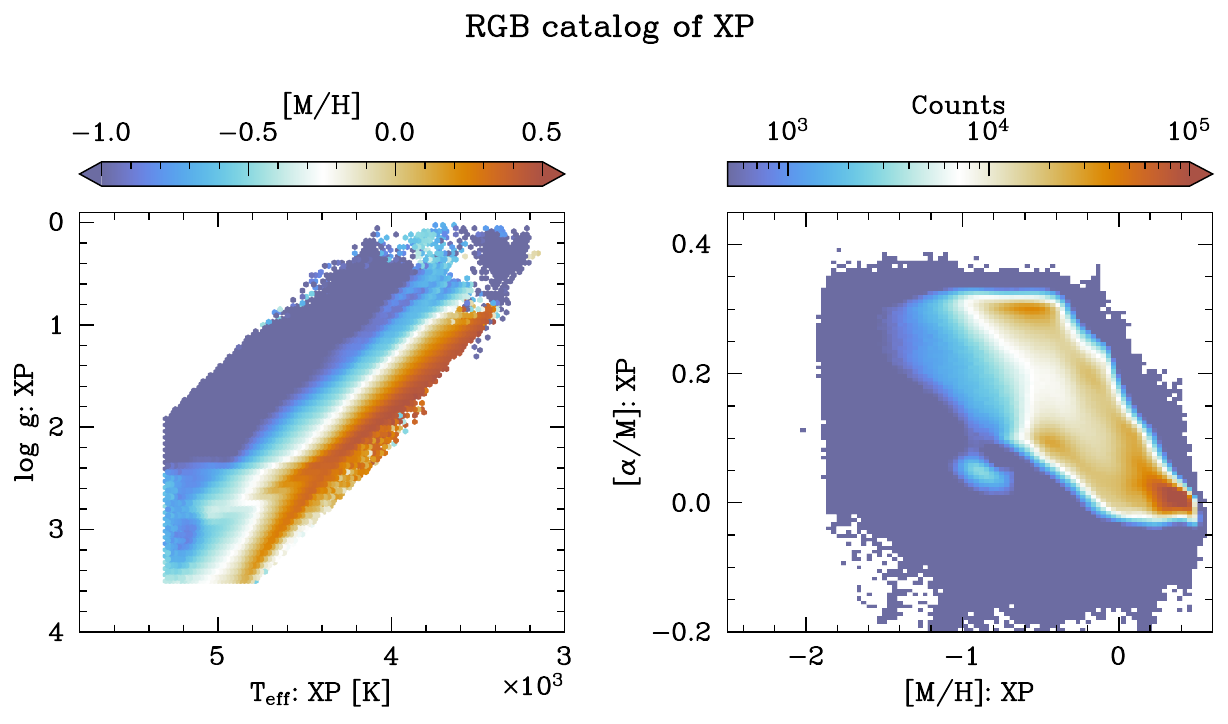}
    \caption{Summary of the stellar labels derived via \xp~ for the $\sim$ 23 million members of the RGB catalog (see Fig.~\ref{fig:hrdcut} 
    Left: \teff--\logg\ diagram color coded by \moh.
    Right: \moh-\aom\ abundance diagnostic diagram color coded by logarithmic density.}
    \label{fig:cat}
\end{figure*}

\begin{figure*}
    \centering
    \includegraphics[width=\linewidth]{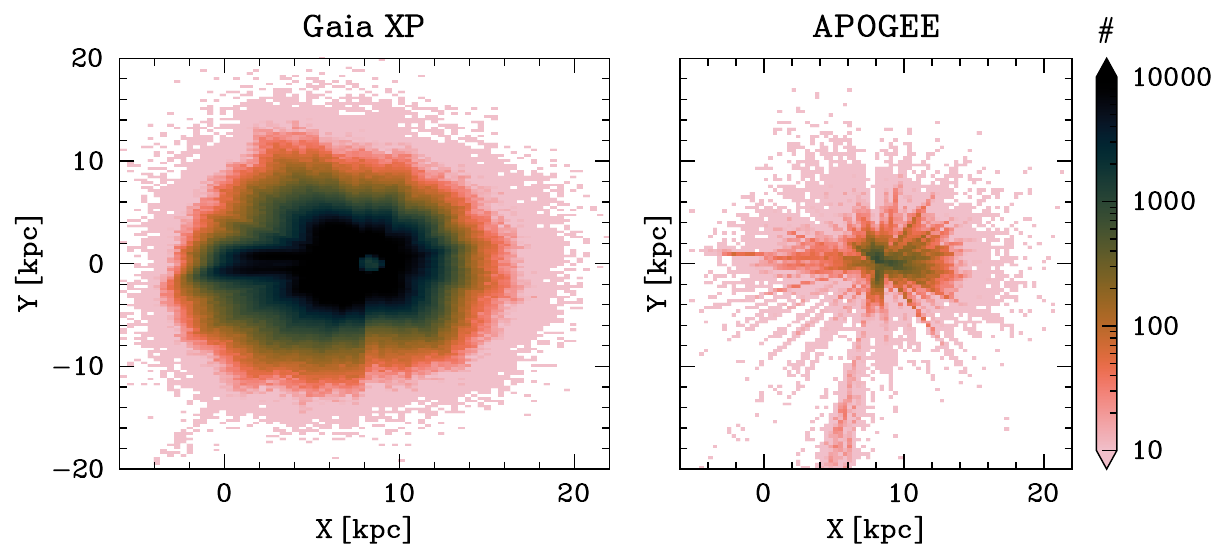}
    \caption{ The left panel shows the 2D distribution of the median galactocentric Cartesian X position {\it vs} Cartesian Y position for \gaia\ XP. 
    On the right panel, a similar 2D distribution is displayed for \apogee\ data. 
    By utilizing the sample provided by XP, we can assess whether the larger and all-sky \gaia\ XP sample covers the same physical extent as the \apogee\ data.
    % \HWR{Please tell the reader what to make of this: bigger sample, same physical extent, ??}
    }
    \label{fig:XYcompare}
\end{figure*}

We now present stellar labels (\teff, \logg, \moh, and \aom) derived in this study, and summarized a catalog (Table~\ref{tab:catalog}).
We first use the ADQL query as presented in Appendix~\ref{app:220m} to obtain $\sim 220$ million stars with available XP data.
% then we apply the trained \xp\, to predict labels for all the $\sim 220$ million stars. 
Then we employ one of the trained models from the 2-fold training set as the inference model. The labels and their corresponding uncertainties generated by the XP decoder serve as the estimated labels provided by XP.
Additionally, we exclude hot stars by selecting those with $G_{\rm BP} - G_{\rm RP} > 0.8$.
Second, we follow the conditions 1-5 described in subsection~\ref{subsec:trdata} to select physically reasonable stellar labels. 
Third, we apply the following conditions to select a high-quality sample of approximately 28 million RGB stars, defined by:
\begin{enumerate}
    \item \teff/\eteff$>10$;
    \item \logg/\elogg$>10$;
    \item \eaom$<0.5$;
    \item \emoh$<1$;
\end{enumerate}

We note that such S/N-based sample cuts produce clean  and easy to work with sub-samples, but may cause some difficulties if these cuts become part of a modelling selection function \citep{Rix2021}.
As the last steps we make simple cuts in the HR (or 'Kiel') diagram, as shown in Fig.~\ref{fig:hrdcut}, to distill a relatively pure RGB sample.
We adopt a pseudo-luminosity ($L_{\rm pseudo}$) \citep{Anderson2018} to select giant stars in HRD, instead of absolute magnitude $M_G$ to avoid losing sample members with negative parallax measurements (about $\sim2$\%):
\begin{equation}
    L_{\rm pseudo} = \varpi \cdot 10^{G/5} < 10^{(-0.003 T_{\rm eff} + 19)/5+10},
\end{equation}
where $L_{\rm pseudo}$ is a scaling value of luminosity that equals to $10^{M_G/5+2}$.
This cut is effectively equal to $M_G > -0.003 T_{\rm eff} + 19$ for high-quality parallax measurements.
This final cut leaves $\sim 23$ million stars, which we illustrate in Fig.~\ref{fig:cat}.

We deem this RGB sample to be most suitable for astrophysical analyses where \aom~ may play a role, and we present it as a catalog in Table.\ref{tab:publishcat}. It can be accessed via \url{https://zenodo.org/record/8002699}.
Figure~\ref{fig:XYcompare} illustrates the larger coverage of the galactocentric Cartesian X-Y plane by the RGB catalog compared to \apogee. 
The all-sky nature and sample size of our RGB catalog from \gaia\ XP enhances our capabilities in Galactic Archaeology, particularly when it comes to obtaining alpha abundance estimates.

\begin{table*}[htbp]\label{tab:publishcat}
    \centering
    \caption{Table descriptions for $\sim 23$ million RGB stars predicted by \xp.}
    \label{tab:catalog}
    \begin{tabular}{ll}
        \hline
        \textbf{Column Name} & \textbf{Description (Units)} \\
        \hline
        source\_id & Unique identifier for star \\
        ra & Right Ascension (deg) \\
        dec & Declination (deg) \\
        teff\_xp & Effective temperature (K) from \xp\\
        logg\_xp & Surface gravity (dex) from \xp\\
        moh\_xp & Metallicity (dex) from \xp\\
        aom\_xp & $\alpha$-element abundance (dex) from \xp\\
        e\_teff\_xp & Error in effective temperature (K) from \xp\\
        e\_logg\_xp & Error in surface gravity (dex) from \xp\\
        e\_moh\_xp & Error in metallicity (dex) from \xp\\
        e\_aom\_xp & Error in $\alpha$-element abundance (dex) from \xp\\
        snr\_rp & Signal-to-noise ratio in $G_{RP}$\\
        l & Galactic longitude (deg)\\
        b & Galactic latitude (deg)\\
        parallax & Parallax (mas)\\
        parallax\_error & Parallax uncertainty (mas)\\
        pmra & Proper motion in RA (mas/yr)\\
        pmra\_error & Proper motion uncertainty in RA (mas/yr)\\
        pmdec & Proper motion in Dec (mas/yr)\\
        pmdec\_error & Proper motion uncertainty in Dec (mas/yr)\\
        ruwe & Renormalized unit weight error\\
        phot\_g\_mean\_mag & Mean apparent magnitude in $G$ band\\
        phot\_bp\_mean\_mag & Mean apparent magnitude in $BP$ band\\
        phot\_rp\_mean\_mag & Mean apparent magnitude in $RP$ band\\
        bp\_rp & Color index\\
        radial\_velocity & Radial velocity (km s$^{-1}$)\\
        radial\_velocity\_error & Radial velocity uncertainty (km s$^{-1}$)\\
        \hline
    \end{tabular}
\end{table*}

%%%%%%%%%%%%%%%%%%%%%%%%%%%%%%%%%%%%%%%%%%%%%%%%%%%%%%%%%
%
%
%
%    Section: validation        
%
%
%
%%%%%%%%%%%%%%%%%%%%%%%%%%%%%%%%%%%%%%%%%%%%%%%%%%%%%%%%%
\subsection{Accuracy evaluated by external stellar labels}

\subsubsection{Comparison with A23}

\begin{figure*}[htbp]
    \centering
    \includegraphics[width=0.99\linewidth]{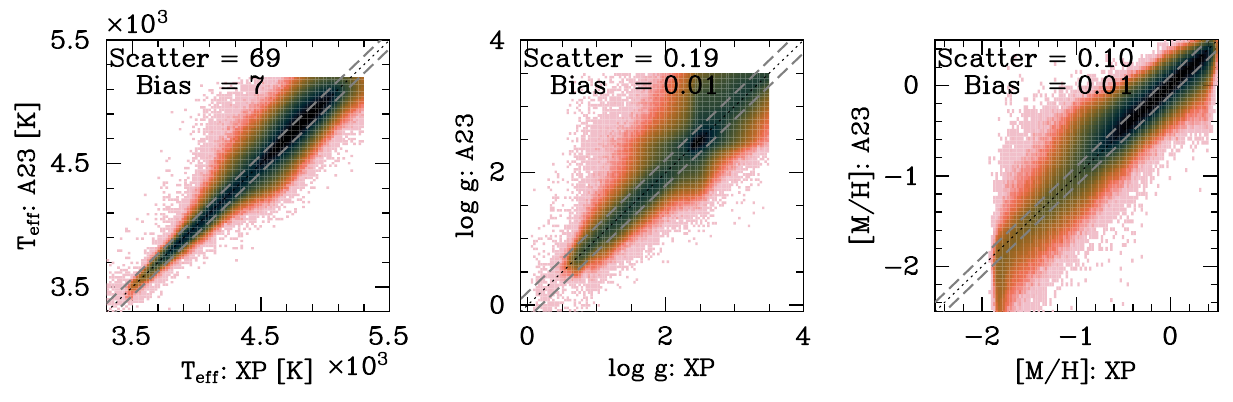}
    \includegraphics[width=0.99\linewidth]{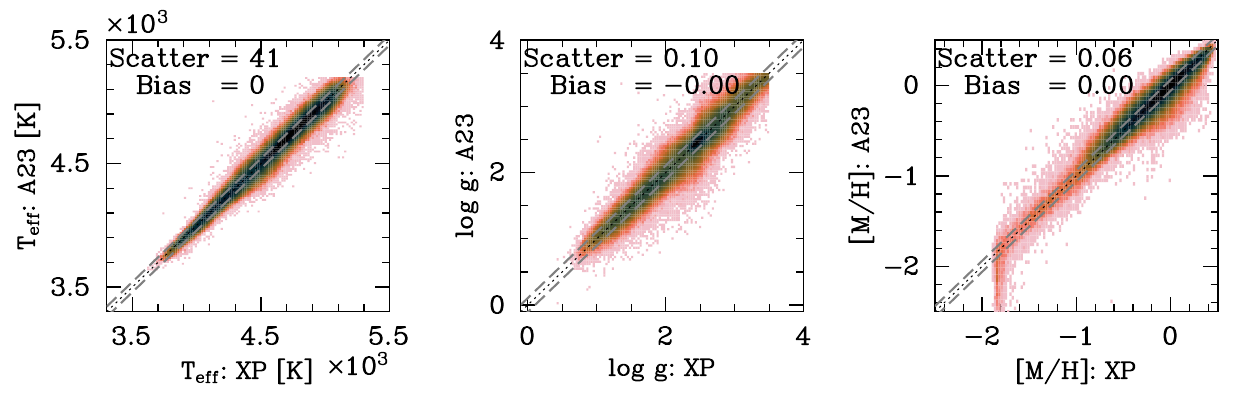}
    \caption{Comparison of the stellar labels \teff, \logg, and \moh\ (left to right), derived from essentially the same XP data but with two different approaches \xp~ and XGBoost as implemented by \citetalias{2023arXiv230202611A} \citep{2023arXiv230202611A}. (Note that \citetalias{2023arXiv230202611A} used also band-pass magnitudes derived from  spectra  and additional external photometry, such as WISE.)
    The figures are color-coded by the logarithmic sample number density.
    The first row of figures represents the comparison for the entire cross-matched RGB sample, consisting of 10,825,736 stars. 
    In the second row, the comparison is restricted to the subset of stars with a signal-to-noise ratio (S/N) of RP coefficients larger than 1,000, comprising 917,921 stars.}
    \label{fig:rene23}
\end{figure*}

% We compared the labels provided by \xp\ with those derived by \cite{2023arXiv230202611A} (A23) for 13,300,628 stars in the RGB catalog.
% After cross-matching this catalog with our \xp\ catalog, we obtained a sample of 11,668,067 stars.

We compared the labels provided by \xp\ with those derived by \cite{2023arXiv230202611A} (\citetalias{2023arXiv230202611A}) for 13,300,628 stars in the RGB catalog. 
To start, we note that both \xp\ and \citetalias{2023arXiv230202611A} utilize XP spectra and train on APOGEE data. However, they use quite different approaches, as \citetalias{2023arXiv230202611A} employs a direct discriminative machine learning trained on APOGEE. 
As shown in Fig.~\ref{fig:rene23}, the scatter between \xp\ and \citetalias{2023arXiv230202611A} for \teff, \logg, and \moh\ were found to be 73~K, 0.2~dex, and 0.1~dex, respectively, for all the cross-matched stars, with nearly no biases for the labels from the two catalogs, only 7~K, 0.02~dex, and 0~dex. 
The higher quality labels given by \xp, with \eteff\ smaller than 50~K, \elogg\ less than 0.1~dex, and \emoh\ $<0.1$~dex, were also shown in Fig.\ref{fig:rene23}.  
We find that the scatter in \teff, \logg, and \moh decreases to 45~K, 0.12~dex, and 0.07~dex. 
The \moh\textsubscript{XP} are systametically higher from \citetalias{2023arXiv230202611A} for metal-poor regime (\moh$<-2$), this is due to the limitation of our training sample, the majority of \moh\ from the \apogee\ label as traininig dataset is larger than -2~dex, 
but \citetalias{2023arXiv230202611A} add very-metal-poor star training sample [Li23] to benefit the estimation of the metal-poor regime.

To calculate the combined uncertainty when comparing the two catalogs, we can simply use $   \delta_\mathrm{comb} = \sqrt{\delta_\mathrm{1}^2 + \delta_\mathrm{2}^2},$
where $\delta_\mathrm{1}$ and $\delta_\mathrm{2}$ are the uncertainties of the two catalogs.
Given the typical error of 50~K, 0.12~dex, and 0.07~dex for \teff, \logg, and \moh for \xp, 50~K, 0.08~dex and~0.1 dex for \citetalias{2023arXiv230202611A}, the combined uncertainties for three labels ( \teff, \logg, and \moh) are 70~K, 0.14~dex and 0.12~dex, assuming the errors are uncorrelated. 
The rough estimates are consistent with the comparison as displayed in Fig.~\ref{fig:rene23} for \teff\ and \moh, the scatter of \logg\ between \citetalias{2023arXiv230202611A} and \xp\ are larger 0.06~dex.
After the quality cut of \xp, the scatters are nearly the same as shown in the self-validation value reported in subsection~\ref{subsec:error}.

We find that 88\% of the \citetalias{2023arXiv230202611A} stars are contained in our catalog,
which implies that much of the \emph{RGB} sample of \citetalias{2023arXiv230202611A} also overlaps. 
However, it should be noted that our catalog is based only on \gaia\ DR3 data, while the parent catalog of \citetalias{2023arXiv230202611A} includes Gaia DR2, 2MASS, and ALLWISE surveys, resulting in a smaller parent catalog size compared to the \xp\ catalog of approximately 23 million stars, mostly caused by ALLWISE incompleteness.
It is plausible that the main difference in catalog content between \citetalias{2023arXiv230202611A} and this work is attributable to the S/N cuts we used for selecting stars. 
% A23 do not apply quality cut, but the \xp-catalog has a more stringent error cut as described before, it may exclude more stars selected by A23 and have a smaller containment.

\subsubsection{Comparision with LAMOST-LRS}

\begin{figure*}[htbp]
    \centering
    \includegraphics[width=\linewidth]{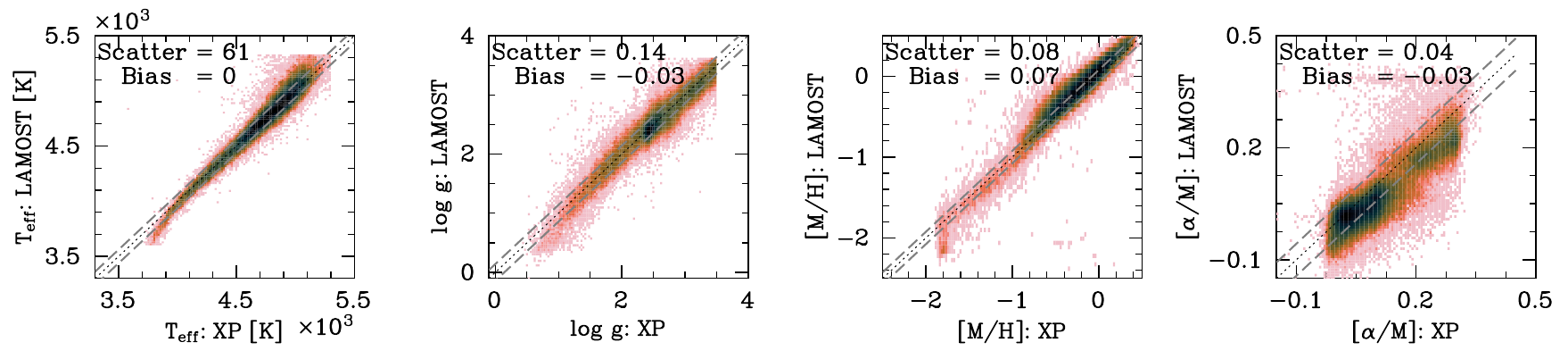}
    \includegraphics[width=\linewidth]{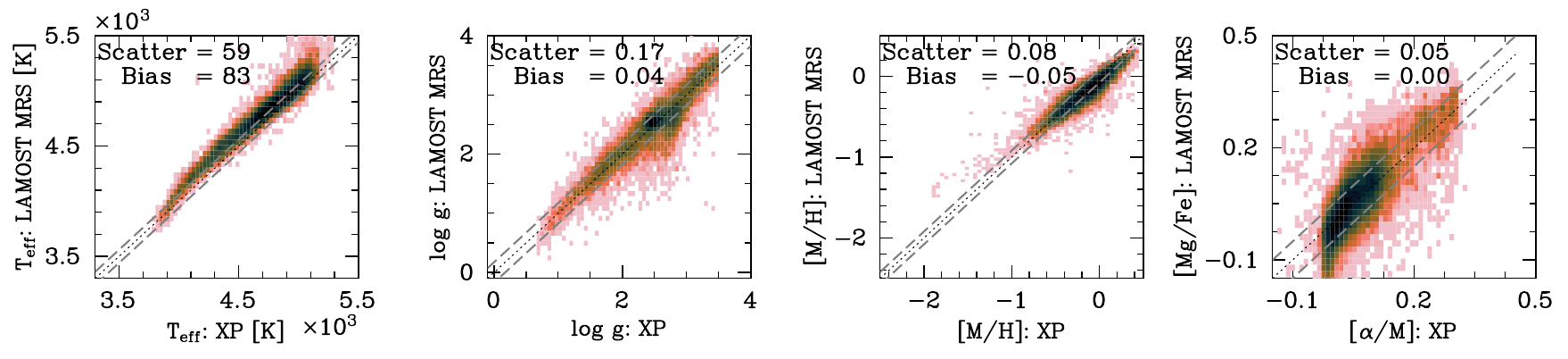}
    \includegraphics[width=\linewidth]{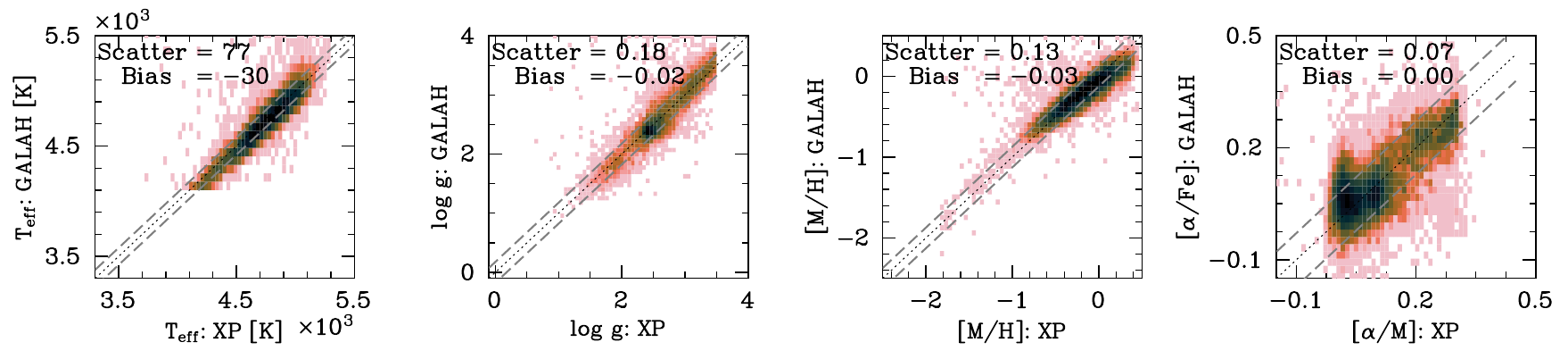}
    \caption{
     Comparison between \xp-derived labels and those from three different ground-based survey datasets: LAMOST-LRS, LAMOST-MRS, and GALAH. Each row corresponds to a specific parameter: \teff, \logg, \moh, and \aom.
     The top panel represents the comparison with \gaia\ XP (\xp) and LAMOST-LRS (SLAM), color-coded by the number density. 
     The middle panel showcases the comparison with LAMOST-MRS dataset (Cycle-StarNet), and the bottom panel depicts the comparison with GALAH DR3.
     In each panel, the scatter and the median bias for each stellar label are marked.
    The black dotted line represents the one-to-one line, reflecting perfect agreement between the surveys. 
    The pairs of grey dashed lines in the plot represent the deviation of one sigma in the difference between the compared labels.
    }
    % \HWR{(of what? the scatter of the 1sigma of the bias?)}, 
    \label{fig:lamost}
\end{figure*}

For an independent validation with R$\gtrsim 2000$ spectroscopy, we use datasets beyond the \apogee\ survey to verify the performance of our results. 
In Figure \ref{fig:lamost}, we compare the labels derived by \xp\ for red giant branch (RGB) stars from LAMOST Low-Resolution  ($\mathcal{R} \sim 1800$) Spectroscopic Survey (LRS) DR5 \citep{zhang2020a} using the Stellar LAbel Machine (SLAM). 
The labels of LAMOST RGB stars are trained by transfering approximately 90,000 common stars between LAMOST DR~5 and APOGEE DR~15.
We first select stars with S/N$>100$ (in SDSS $g$-band) from the SLAM catalog\footnote{\url{https://github.com/hypergravity/paperdata}}, with the flag set to K giant, resulting in 142,608 stars.
The S/N cut ensures that the errors of the SLAM labels are less than $\sim$ 50~K, 0.1~dex, 0.037~dex, and 0.026~dex for \teff, \logg, \moh, and \aom, respectively \citep{zhang2020a}. 
We then cross-matched the selected LAMOST K giant stars with our catalog, resulting in 116,061 common stars.
The comparisons between \xp~ labels from \gaia\ XP and SLAM labels from LAMOST are shown in Fig. \ref{fig:lamost}. 
Generally, the differences between the \xp-labels and LAMOST are similar to the results of the validation result shown in Fig. \ref{fig:compare4d}. 
For \teff, the difference is larger at the edge of the label range, i.e., \teff~$<4200$ and \teff~$>5000$. 
Similarly, the scatter of \logg\ is higher in the range of \logg $< 1.5$. 
One reason for this could be that the K-giant stars from the LAMOST catalog are selected using a hard cut on \teff\ and \logg\ (Fig. 6 in \citealt{zhang2020a}), 
which may erroneously include main-sequence stars mixed with RGBs.
To validate our conjecture, we further cut the sample with \teff\ difference less than 200K, the number of outliers is 2,006 outliers. 
We find that the number of stars with a large \moh\ difference from LAMOST decreased after the cut. 
Our comparison with LAMOST illustrates the remarkable consistency of the \xp\ labels with the literature, even outside of \apogee\ observation.
However, we should note that both \xp\ and SLAM are limited in its truncation in the \apogee\ training dataset. 

For \moh\ and \aom, we find some spurious differences in the metal-poor regime (\moh$<-1$). 
We suspect that the \xp-derived \moh\ suffers from significant errors for metal-poor stars, which is consistent with our findings in the validation results shown in Fig. \ref{fig:cat}.
Additionally, we find large discrepancies for \aom$>0.2$, but that might be a result of large errors in \moh\ for metal-poor stars, where \moh\ and \aom\ are strongly degenerate in this challenging regime for abundance estimation.
Although the scatters of \moh\ between \xp\ and SLAM are only 0.07~dex, however, there is a 0.06\,dex bias between \xp\ and SLAM.
The bias might come from the updated pipleline between APOGEE DR ~17 and APGOEE DR~14.
We found similar bias for \aom, the bias between \xp\ and SLAM is -0.03~dex. 

\subsubsection{Comparison with LAMOST-MRS}

We further compare our results with \textsl{Cycle-StarNet} from \cite{Wang2023b}, which utilized MARCS model atmospheric theoretical synthetic spectra combined with a domain-adaptation method to estimate the fundamental stellar parameters (\teff, \logg, [Fe/H]) and 11 chemical abundances for 1.38 million stars from the Medium-Resolution ($\mathcal{R} \sim 6500$) Spectroscopic Survey (MRS) in LAMOST-II DR8\footnote{\url{https://nadc.china-vo.org/res/r101242}}.

To perform our comparison, we cross-match our RGB catalog with the dataset from \cite{Wang2023b} and obtain a sample of 301,478 stars. 
We then apply specific selection criteria based on the flags provided by \cite{Wang2023b}, namely {\tt\string Flag\_Teff} = 0, {\tt\string Flag\_logg} = 0, {\tt\string Flag\_FeH} = 0, {\tt\string Flag\_MgFe} = 0, {\tt\string SN\_blue} > 100, {\tt\string SN\_red} > 100, and constrain the errors of [Fe/H] and [Mg/Fe] to be smaller than 0.05 dex and 0.04 dex, respectively.
After applying these data quality cuts, we obtain a final sample of 40,035 stars for our comparative analysis as shown in Fig. \ref{fig:lamost}.

As depicted in Figure \ref{fig:lamost}, the comparison results with LAMOST MRS exhibit similarities to those obtained from LAMOST LRS, despite the application of different methods and spectra. 
Regarding the effective temperature (\teff), the scatter shows a marginal difference, attributed to a slightly larger bias ranging from 0~K to 83~K. 
In terms of \logg, the scatter is slightly larger, measuring up to 0.17 dex. 
For metallicity (\moh), the scatter remains comparable to the comparison with LAMOST MRS; however, there appears to be a smaller representation of metal-poor stars (\moh$<-1$) in the LAMOST MRS sample. 
Since \cite{Wang2023b} does not provide an overall $\alpha$ abundance, we compare the \xp-derived \aom\ with [Mg/Fe] as a reference. 
We find that the scatter is similar to that obtained for SLAM. 
In summary, we find good agreement between \teff, \logg, \moh, and \aom\ derived from \xp\ and those obtained from LAMOST MRS.

\subsubsection{Comparision with GALAH}

We further conducted a comparison with the results obtained by \cite{Buder2021}. 
The GALAH DR3 consists of 768,423 high-resolution (R $\sim$ 28,000) optical spectra obtained from 342,682 stars. 
The stellar parameters in GALAH DR3 were estimated using the Spectroscopy Made Easy (SME) model-driven approach in combination with 1D MARCS model atmospheres. 
Additionally, \cite{Buder2021} incorporated astrometric information from \gaia\ DR2 and photometric data from 2MASS to mitigate spectroscopic degeneracies, accounting for LTE/non-LTE effects in their computations.

To perform a comparative analysis, we cross-matched our results with GALAH DR3 and identified 13,504 stars with corresponding stellar parameters. 
Quality flags were applied to ensure the reliability of the the GALAH labels, including {\tt\string flag\_sp}=0, {\tt\string flag\_fe\_h}=0, {\tt\string flag\_guess}=0, and {\tt\string red\_flag}=0. 
Moreover, we imposed constraints on the errors of \teff, \logg, [Fe/H], and [$\alpha$/Fe] given by GALAH DR3, setting them to be smaller than 200~K, 0.25~dex, 0.2~dex, and 0.05 ~dex, respectively.
After applying these quality criteria, we obtained a subset of 13,504 stars for the comparative analysis as displayed in Fig. \ref{fig:lamost}.

The comparison between \xp\ and GALAH reveals a consistent pattern, although the level of consistency is relatively weaker compared to A13, LAMOST LRS, and LAMOST MRS. 
The scatter values for \teff, \logg, \moh, and \aom\ amount to 77 K, 0.18 dex, 0.13 dex, and 0.07 dex, respectively. 
In terms of bias, the four labels show a slight deviation, with a bias of -30 K for effective temperature (\teff).

The comparative analysis of various surveys' similarity in labeling is presented in \cite{Wang2023b}. 
LAMOST MRS and APOGEE exhibit the highest level of consistency in their labels, followed by GALAH. 
The SLAM labels (derived from LAMOST LRS) are trained using labels from APOGEE DR14, hence the expected consistency between them.

In conclusion, we performed independent validation by comparing the labels derived by \xp\ with the LAMOST Low-Resolution Spectroscopic Survey (LRS), the Medium-Resolution Spectroscopic Survey (MRS) datasets, and GALAH DR3. 
The comparison results demonstrate the remarkable consistency of the \xp\ labels with the literature, indicating the accuracy and reliability of the \xp\ model in providing stellar labels. 
However, it is important to note that both \xp\ have limitations in their training datasets, which are based on the truncation of the \apogee\ dataset.

\subsection{The accuracy of metallicity assessed from clusters}

\begin{figure*}[htbp]
    \centering
    \includegraphics[width=\linewidth]{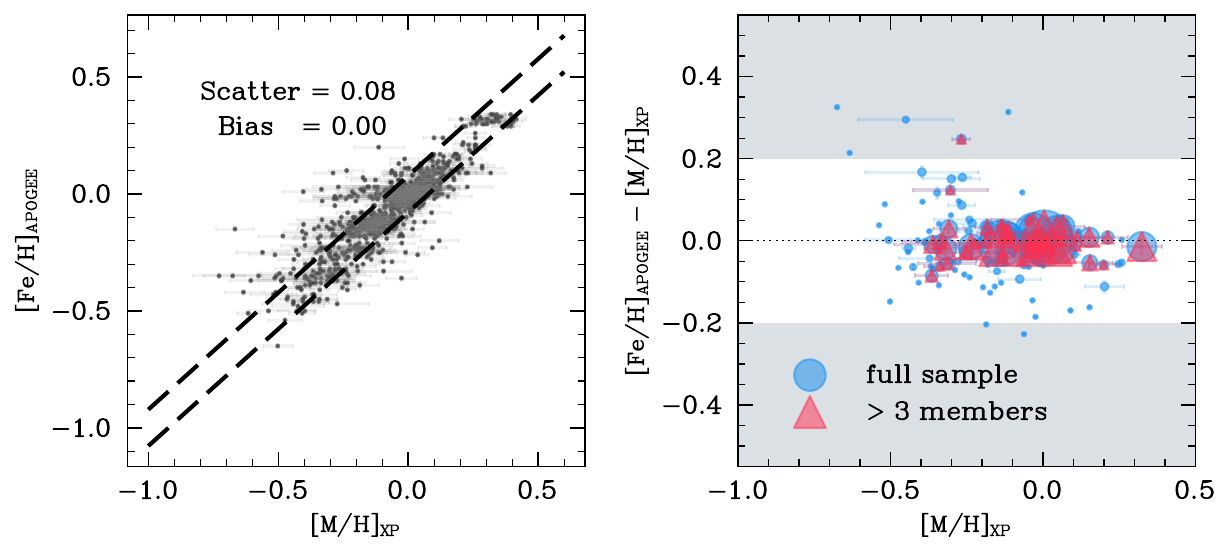}
    \caption{
    % \HWR{This Figure needs re-writing. APOGEE is mentioned throughout, but never appears on any of the X- or Y-axes. I was confused. Also: "All sample" --> "full sample" and "Members >3" --> ">3 members".}
    Comparison between literature [Fe/H], \xp, and \apogee\ DR16 \moh\ for open clusters. 
    We compare the difference between \xp\ \moh\ and [Fe/H] in  \cite{Donor2020}. 
    The blue circles represent the \xp\ \moh, and the red triangles represent the \apogee\ \moh, with 1$\sigma$ as the error bar.
    The metallicity difference $-0.2<\Delta$\moh$<0.2$ is shown by the gray shaded areas.
    All cluster samples cross-matched with literature are shown in the left panel, and 12 selected clusters with more than five identified members in \apogee\ are shown in the right panel.
    The marker size indicates the number of members in each cluster.
    }
    \label{fig:cluster}
\end{figure*}

To assess the accuracy of the abundances estimated in our work, we explore the abundance derived with \xp\ for stars in open clusters. 
Stars in open clusters serve as benchmark stars, which are due to the approximately chemically homogeneous. 
In Figure \ref{fig:cluster}, we compare \moh\ from \xp\ with the literature values for 67 known open clusters \citep{Donor2020}. 
% The open cluster members are identified in \cite{Donor2020}, we select the member with . 
% For the monolithic [M/H] of the cluster, we take the median of all members of each cluster.
The difference between the \xp\ and the literature values, as well as the APOGEE \moh\ compared to the literature values, are shown in Fig.\ref{fig:cluster}. 
We find that although the deviations of the estimates from \xp\ are larger than those of \apogee, 66 out of 67 clusters have \xp\ \moh within 0.2 dex. 
However, the selected 12 clusters with more than 3 member stars, the deviation of the \xp\ derived \moh\ from the literature values is within 0.07 dex as shown in Figure \ref{fig:cluster}.
There is no metallicity dependence for the deviations of estimates from \xp. 
Testing \xp\ on open clusters shows that the error of the \moh\ estimate is mainly the random error when compared to \apogee\ abundances.

%%%%%%%%%%%%%%%%%%%%%%%%%%%%%%%%%%%%%%%%%%%%%%%%%%%%%%%%%
%
%
%
%    Section: $[\alpha/M]$-abundance        
%
%
%
%%%%%%%%%%%%%%%%%%%%%%%%%%%%%%%%%%%%%%%%%%%%%%%%%%%%%%%%%

\subsection{Verification of the \aom-abundances}

The task of determining the $[\alpha/M]$-abundance is challenging for low resolution spectra like \gaia\ XP. This is due to the potential degeneracy between the \aom -abundance and other parameters, such as the ``metallicity'' (\moh ), as shown in studies by \cite{Ting2017a,Gavel2021}.
In this section, we evaluate the precision and accuracy of our $[\alpha/M]$-abundance estimates, thereby validating the \aom\, derived from \xp.  
We conduct three main tests to validate the $[\alpha/M]$-abundance prediction. These tests in particular include circumstances where we know how completely independent properties of the stars (such as their positions or velocities) correlate with $[\alpha/M]$; we then check whether these correlations are seen at the expected level.

First, we examine how well \xp\, can distinguish between different chemical components of the disk, specifically the so-called low-$\alpha$ and high-$\alpha$ disk, within the same range of \moh. 
This analysis is discussed in detail in subsection~\ref{subsubsec:disk}.
Then we explore the relationship between the orbit dynamics and \aom-abundance for disk stars : do high-$\alpha$ stars (at a given \moh ) form a hotter disk that low-$\alpha$ stars of the same \moh. 
For a more detailed discussion, please refer to Subsection \ref{subsubsec:orbits}.
While our comparison with LAMOST and GALAH data demonstrates agreement between \xp-derived $[\alpha/M]$-abundance and these surveys, it is important to note that the majority of LAMOST and GALAH samples consist of disk stars with \emoh$>-0.8$, leaving the question open, whether our \aom~ estimates can differentiayte stellar population also in the metal-poor regime.
To offer further independent validation, we specifically chose stars from halo substructures and the Large Magellanic Cloud (LMC), which exhibit distinct star formation histories in comparison to the disk stars within the Milky Way. 
The validation process and the corresponding results will be thoroughly discussed in Subsection~\ref{subsubsec:ges}.

\subsubsection{Assessing accuracy through the bimodality of disk Stars}\label{subsubsec:disk}

The well-established bimodality of $[\alpha/M]$-abundance in disk stars serves as the initial validation of our ability to separate chemically  low-$\alpha$ (or ``thin disk'') from  high-$\alpha$ (or, ``thick disk'') stars. 
To accomplish this, we employ a two-fold validation approach using \xp\ labels. Specifically, we compare a selected sample of disk stars with $-0.9<$\moh$<0$ against the \apogee\ training labels to evaluate the effectiveness of differentiating between low-alpha and high-alpha disk populations, the details can be found in Appendix~\ref{app:disk}.

As shown in Table~\ref{tab:precision_recall}, for the low-alpha disk, we find that 97\% of the XP-identified group corresponds to the low-alpha class according to \apogee\ labels, while 96\% of the entire low-alpha disk sample identified by \apogee\ is correctly classified by XP. Similarly, for the high-alpha disk, 93\% of the XP-identified group represents the high-alpha class, and 94\% of the complete high-alpha sample identified by \apogee\ is accurately recognized by XP. 
These validation results demonstrate that the \xp-\aom\ exhibits a high level of precision and recall in distinguishing between low-alpha and high-alpha disk populations. Further details can be found in Appendix~\ref{app:disk}.

\begin{table}[htbp]
  \centering
  \caption{Precision and recall for low-\aom~ disk and high-\aom~ disk}
  \label{tab:precision_recall}
  \begin{tabular}{@{}rll@{}}
    \toprule
    Class & Precision & Recall \\
    \midrule
    low-\aom~  disk & 0.97 & 0.96 \\
    high-\aom~ disk & 0.93 & 0.94 \\
    \bottomrule
  \end{tabular}
\end{table}

\subsubsection{\aom~ validation via orbit dynamics}\label{subsubsec:orbits}

\begin{figure*}[htbp]
    \centering
    \includegraphics[width=\linewidth]{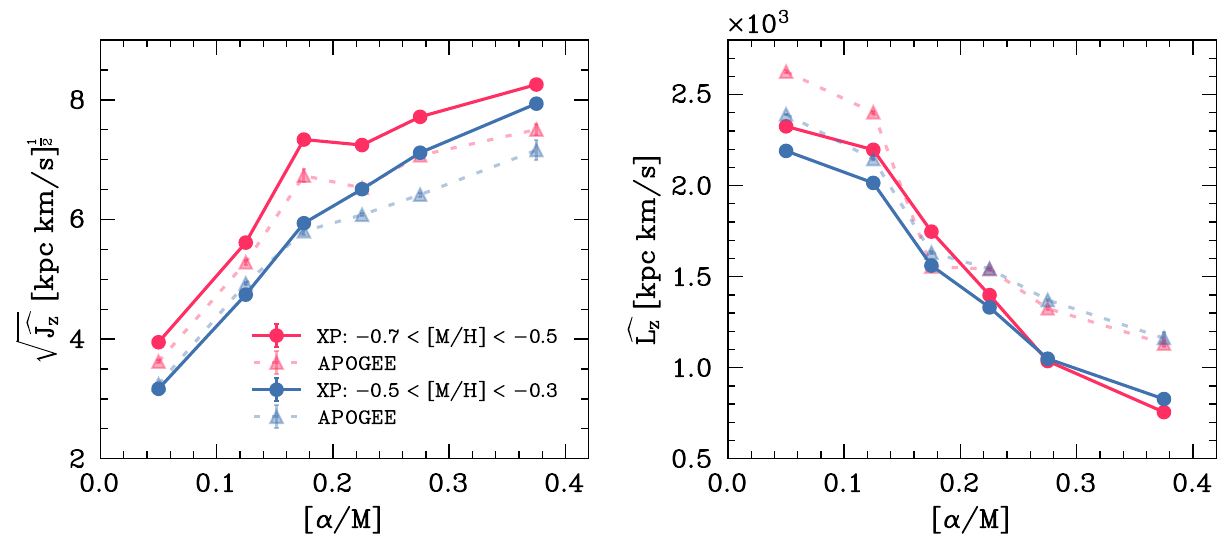}
    \caption{Astrophysical validation of \xp~ 's \aom~ determinations, based on the fact that -- at a given \moh~-- the $\alpha$-enhanced (or 'thick') disk has more vertical motions \citep[e.g.][]{Bovy2012}. The left panel illustrates the relationship between the mean root-square vertical action $J_z$ (quantifying the vertical kinematics) and \xp-derived \aom~ in two narrow bins of \moh,  $-0.7<$\moh$<-0.5$ (red) and $-0.5<$\moh$<-0.3$ (blue). 
    The solid line with circles shows the result of \xp-labels, while the dashed lines with triangles represent those from \apogee\ for reference. 
    The \xp~ results show the expected trend that agrees quite closely with that seen in \apogee data; subtle offsets may simply reflect the different spatial selection function of the sample.
    The right panel shows a related astrophysical validation test, based on the known fact that -- again at a given \moh~-- the $\alpha$-enhanced disk is much more centrally concentrated \citep[e.g.][]{Bovy2016}. 
    The panel displays the mean angular momentum $L_z$ as a function of \aom~in the same two \moh -bins, again showing the expected trend and good agreement with \apogee .}
    \label{fig:aom_jz_lz}
\end{figure*}

\begin{figure*}[htbp]
    \centering
    \includegraphics[width=0.49\linewidth]{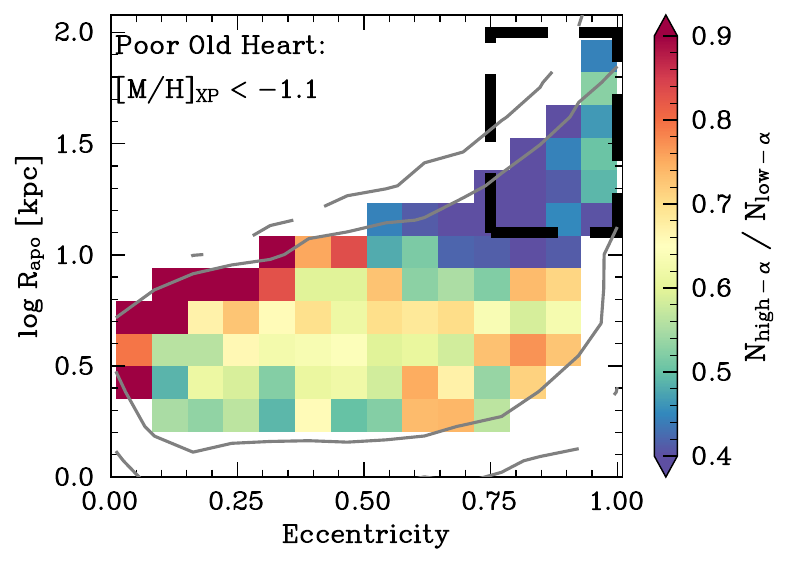}
    \includegraphics[width=0.48\linewidth]{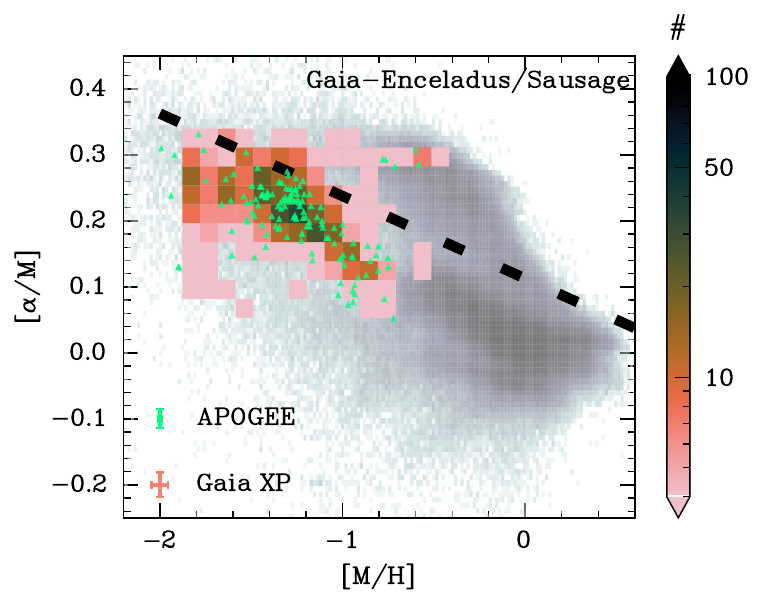}
    \includegraphics[width=0.55\linewidth]{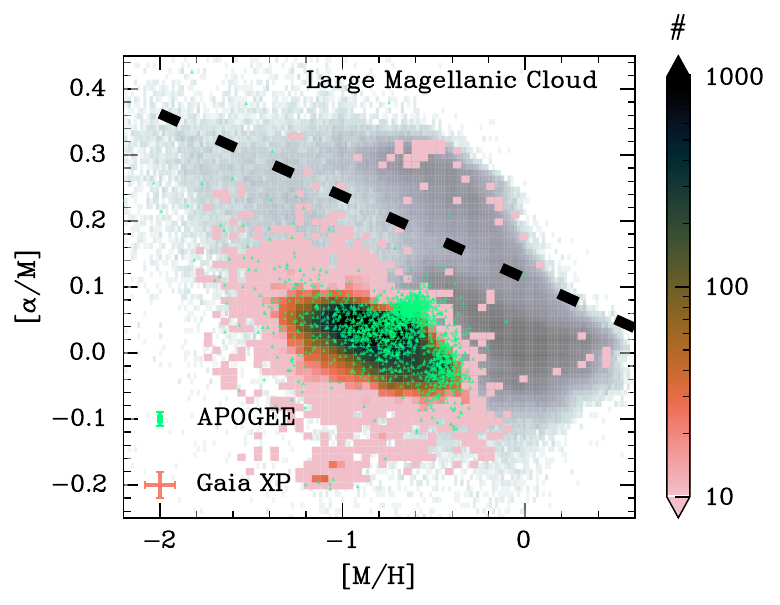}
    \caption{Three astrophysical validation tests for the quality of the \xp -based \aom~ estimates in the metal-poor regime. All three validations are based on the idea that in some regimes the spatial or kinematic selection of stars leads to clear (externally derived) expectations for $p($\aom ) in a given \moh -regime.
    {\bf{\textit{Top left}}}: Validation using the \emph{Poor Old Heart} of he Milky Way \citep{Rix2022}, showing the fraction of high-[$\alpha$/M]\textsubscript{XP} stars as a function of $R_{\rm apo}$ and eccentricity for stars with [M/H]\textsubscript{XP} $<-1.1$. 
    At high eccentricities ($>$0.75)and large $R_{\rm apo}$ ($>10$~kpc) the population should be dominated by the Gaia-Enceladus/Sausage (GSE) population, known to be low-$\alpha$ \citep{Helmi2018,Hasselquist2021}
    % \HWR{~REF?~}
    ; this is exactly what our [$\alpha$/M]\textsubscript{XP} values show.
    {\bf{\textit{Top right}}}: Distribution of stars in the \moh-\aom plane that have been selected (\emph{purely kinematically}) as likely GSE members, expected to lie below the (black dashed) high-$\alpha$~\emph{vs.}~low-$\alpha$ dividing line in the diagram. The stars with \xp -determined \moh~ and \aom ~labels (colored density) lie in the extected position, and also agree with \apogee (green points); this latter coincidence is not a trivial consequence of the training, given that these stars are a tiny and unusual subsample. 
    {\bf{\textit{Bottom}}}: Validation in LMC, using a similar \moh-\aom diagonal diagram as the top left panel. The stars with \xp -determined labels (colored density) lie at low \aom~, as expected for the LMC \citep{Russell1992};
    % \HWR{REF?}; 
    they again agree with the \apogee\ labels for this peculiar subset.
     }
    \label{fig:gse_lmc}
\end{figure*}

From the RGB catalog compiled from \gaia XP, we focused on stars within two fixed [M/H] groups: $-0.7 < \text{[M/H]} < -0.5$ and $-0.5 < \text{[M/H]} < -0.3$. 
We cross-matched these groups with the sample from \cite{Kordopatis2023} to obtain orbital parameters based on \textsl{Gaia} DR3. 
We obtained a total of 960,661 and 1,644,704 stars in the two [M/H] groups, respectively.

Figure \ref{fig:aom_jz_lz} illustrates a clear trend: within the range $-0.8 < \text{[M/H]} < -0.3$, the "high-$\alpha$ disk" is vertically hotter (higher vertical action, $J_z$), has older age, and lower angular momentum ($L_z$), while the "low-$\alpha$ disk" is vertically cooler (lower $J_z$), has younger ages, and higher angular momentum.

For reference, we overlaid the analogous APOGEE results in Figure \ref{fig:aom_jz_lz}. 
In general, \gaia\ XP and APOGEE exhibit similar trends on the \aom-$J_z$ diagram, indicating a consistent relationship between $\alpha$-abundance and vertical motion. 
\gaia\ XP also allows for a distinction between different [M/H] groups. 
On the $\alpha$-abundance - $L_z$ panel, APOGEE consistently yields higher values compared to \gaia\ XP. 
This offset may be attributed to a selection effect, as \gaia\ XP contains a larger number of stars in the inner disk with lower values of angular momentum ($L_z$).

Furthermore, we compared the distributions of $\alpha$-abundance and Galactocentric Cartesian X positions between \gaia\ XP and APOGEE for the group with $-0.5 < \text{[M/H]} < -0.3$. 
We observed that the sampling in \gaia\ XP is more uniform across galactocentric X positions, particularly for stars in the inner disk, which have lower angular momentum ($L_z$). This finding further validates our assumption.

Overall, the results confirm \emph{astrophysically} that our  $\alpha$-abundance estimates from \gaia\ XP are meaningful: they clearly can identify the different dynamical properties of stars with different \aom~ at a given \moh , revealing known structural properties of the disk.

\subsubsection{\aom~ validation using the ``Gaia-Enceladus/Sausage" and LMC stellar population properties}\label{subsubsec:ges}

% Verifying the \aom\ in extreme environments such as Gaia-Enceladus/Sausage (GES) and the Large Magellanic Cloud (LMC) is of significant importance. 
% The examination of GES member stars in the \moh-\aom diagram using \apogee\ DR17 catalog provides a foundation for this verification. 
While our comparison with LAMOST and GALAH data demonstrates agreement between \xp-derived $[\alpha/M]$-abundance and these surveys, it is important to note that the majority of LAMOST and GALAH samples consist of disk stars with [Fe/H] $>-0.8$.
To offer further independent validation, we specifically choose stars from halo substructures and the Large Magellanic Cloud (LMC), which exhibit distinct star formation histories compared to the disk stars within the Milky Way.

% \jd{The Galactic Enceladus/Sausage (GES) represents debris from a merger event with a dwarf galaxy. These stars have very different origins compared to Milky Way disk stars. Therefore, validating the \aom\ values for this peculiar population provides a strong test.}
% \jd{Similarly, the LMC has experienced very different chemical enrichment. 
% With its known low $[\alpha/M]$-abundance patterns, it allows testing the \aom\ estimates for stars much farther away than the disk training sample.
% }
% \jd{By examining stars from GES and LMC, we can evaluate whether the \xp-\aom\ values effectively capture the expected chemical signatures, rather than relying on relationships only valid for disk stars. 
% These peculiar stellar populations pose unique challenges for abundance determination.}

Verifying the accuracy of \aom\ measurements in extreme environments, such as Gaia-Enceladus/Sausage (GES) and the Large Magellanic Cloud (LMC), holds significant importance. 
These regions are unique testing grounds for our \aom~ estimates as they have distinct stellar populations from the disk, which dominates our training set. 
GES stars are uncommon halo stars that are beyond the range of thin and thick disk stars, which dominate the training sample. 
Similarly, LMC stars are positioned at significantly greater distances than most training examples. Moreover, general associations observed in typical Milky Way stars, such as a tendency for farther stars to exhibit a higher \aom, cannot be presumed to be applicable to these extragalactic populations. 
The GSE and LMC stars present novel challenges to the model and extend it beyond the standard training data. 
If \xp\, is still able to recover accurate \aom\, values for these outliers, it presents compelling proof that the actual abundance is being measured instead of being only inferred from bulk correlations in the training set. The significance of these tests is pivotal in demonstrating that XP determines elemental abundances and not simply correlates labels.
% Therefore, performing tests on these extreme environments is crucial in ensuring that the \aom\ reflect true stellar properties rather than mimick and propagate spurious correlations or assumptions. 
%To address this, we examine the positions of GES member stars in the \moh-\aom\ diagram using the \apogee\ DR17 catalog as a reference, providing a solid foundation for the verification process.

First, we select high-confidence GES member stars on the basis of their radial orbits. 
Specifically, we select stars with $E_\mathrm{tot} > -1.2 \times 10^5$\,km$^2$\,s$^{-2}$ and $| L_\mathrm{Z} | < 0.5 \times 10^3$\,kpc\,km\,s$^{-1}$, which should confidently exclude the disk and select the most energetic and radial GES stars. 
This selection is motivated by the expectation that GES stars, which were formed in the smaller potential of a former dwarf galaxy, would exhibit lower alpha abundances at fixed [M/H] (e.g., \citealt{Hasselquist2021}.)

We apply further cuts to the sample, requiring \emoh$<0.1$, \eaom$<0.05$, and \logg$<2$ to ensure good quality labels. 
The results for this sample obtained from \xp~ is depicted in Fig.~\ref{fig:gse_lmc}, with \apogee\ labels overplotted as reference. This Figure shows that the vast majority of the stars kinematically selected as GES members, indeed lie below the diagonal line, where high-resolution studies and APOGEE analyses expect them to reside.This represents a confident recognition of the \aom\ values derived by \xp\ for these stars.
We also find that there are a few stars that exhibit chemical characteristics similar to those of the high-alpha disk (above the diagonal line) in both the \apogee\ and \gaia\ XP. 
This phenomenon can be attributed to the dynamical selection of GES stars based on the energy-angular momentum plane, which results in a purity of approximately 24\% and completeness of 41\% as demonstrated in simulations \citep{Carrillo2023}. I.e. thes stars' location in the \aom - moh~ plane may reflect merely the limitations of the kinematic selections. 

% We then cut the sample with \emoh$<0.1$, \eaom$<0.05$, and \logg$<2$ to ensure good labels.
% The results obtained from the \xp, as depicted in Fig.~\ref{fig:gse_lmc}, taking \ap\ labels as a reference.
% We find that the majority of the selected GES stars lie below the diagonal line, about $\sim 87\%$, which indicates that the \xp-derived \aom\ for GES stars are of confidently recognized.

Second, we cross-matched our \xp~ catalog with the catalog of \cite{Rix2022} to analyze the \aom~ abundances of stars  in the \textit{Poor Old Heart} in the Milky Way. 
This cross-matching revealed a common sample of 1,144,026 stars. For the analysis of metal-poor stars in the inner disk, we specifically selected 92,975 stars with \moh$ < -1.1$ based on the results of \cite{Rix2022}. 
We anticipated observing two distinct regimes within this metal-poor sample \citep{Rix2022} . 
The first regime consists of the tighly bound metal-poor stars that have a broad eccentricity distribution ranging from 0.1 to 0.8 (i.e. are approximately isotropic). These stars are members of the \emph{Poor Old Heart}. Our \xp~ estimates show them to be predominately high-\aom~ stars, as expected. The other regime is that of the loosely bound, radially anisotropic stars (with eccentricities  greater than 0.75 and apocenter radii $R_{\textrm{apo}}$ exceeding 10 kpc: these represent the pericenter members of GSE. Consequently, we expect them to be metal-poor, low-\aom~ stars.  Both aspects are affirmed by the top left panel of Figure \ref{fig:gse_lmc}, where we plotted eccentricity versus $R_{\textrm{apo}}$ in Figure \ref{fig:gse_lmc} with  the color-coding representing the \xp-derived [$\alpha$/M]\textsubscript{XP} values. 
Notably, we observed that the stars classified as non-isotropic displayed slightly lower [$\alpha$/M]\textsubscript{XP} values compared to the isotropic stars, as shown in the top-left panel of Figure \ref{fig:gse_lmc}. 
The color bar accompanying the plot indicates the number ratio of high-[$\alpha$/M]\textsubscript{XP} stars to low-[$\alpha$/M]\textsubscript{XP} stars, with high-[$\alpha$/M]\textsubscript{XP} stars defined as those above the diagonal dashed line and low-[$\alpha$/M]\textsubscript{XP} stars below it. 
The results depicted in Figure \ref{fig:gse_lmc} provide additional evidence supporting the reliability of the \xp-derived [$\alpha$/M]\textsubscript{XP} values. 
We observe that the non-isotropic stars, which are likely associated with the debris from the Galactic Enceladus/Sausage (GES) merger, exhibit relatively lower [$\alpha$/M]\textsubscript{XP} values. 
This chemical behavior is consistent with the expected signature of the GES debris, further strengthening the validation of the \aom\textsubscript{XP} in accurately determining $\alpha$-element abundances.

Finally, we validated the \aom\ predictions in the Large Magellanic Cloud (LMC), whihc is a population that is fairly metal-poor with exceptionally low \aom . We cross-matched our sample with the catalog of \cite{Jimenez-Arranz2023}, who employed a supervised Neural Network classifier to distinguish LMC stars from foreground Milky Way stars based on \gaia\ DR3 kinematics data. 
From the common sample of 92,063 stars with high member probability $>0.9$, we selected a subset of 77,277 stars with precise label estimates (\emoh$<0.1$, \eaom$<0.05$, and \logg$<2$). 
We also made sure to include 1,822 stars from \apogee\ DR17. 
The comparison of high-probability LMC member stars with \xp\ \aom\textsubscript{XP} predictions and \apogee\ \aom\ measurements in Figure~\ref{fig:gse_lmc} yields consistent results. 
The \aom\ predictions from \xp\ align with the expected values for the LMC \citep{Russell1992} and are in agreement with the measurements obtained by \apogee. 
This agreement provides further confirmation of the accuracy of the \aom\ predictions, even in challenging validation scenarios.

%%%%%%%%%%%%%%%%%%%%%%%%%%%%%%%%%%%%%%%%%%%%%%%%%%%%%%%%%
%
%
%
%    Section: Discussion        
%
%
%
%%%%%%%%%%%%%%%%%%%%%%%%%%%%%%%%%%%%%%%%%%%%%%%%%%%%%%%%%
\section{Discussion and conclusions}\label{sec:discussion}

\subsection{Caveats}

We have demonstrated how well the low-resolution XP spectra can be used to predict stellar labels, now also including \aom . 
However, our model is based -- inevitably -- on a series of assumptions, and we remind the reader here of the caveats when using our catalog.

\begin{itemize}
    \item 
    Our approach is based on the assumption that the low-resolution spectra are single-star spectra set by only four labels \teff, \logg, \moh, and \aom. We presume stellar rotation, detailed chemical abundances to be negligible, and interstellar extinction to be a nuisance parameter. 
    To break the degeneracy between temperature and extinction, additional infrared photometry, such as 2MASS and ALLWISE \citep{2023arXiv230202611A}, can be incorporated. 
    However, our main objective is to create an RGB catalog solely from \gaia\ data. 
    Consequently, in regions with high extinction, we may encounter challenges in accurately estimating stellar labels.
    Although \cite{Rix2022} has shown that \moh\ estimation remains unbiased even in the presence of significant extinction (e.g., $A_V = 3$), we acknowledge the potential limitations of our stellar label estimates in high-extinction regions.
    In Appendix~\ref{app:further_validation}, we present a comparison with the results of \cite{2023arXiv230202611A} in relatively low-extinction regions characterized by Galactic latitudes $\mid b \mid > 30$, as well as high-extinction regions with $\mid b \mid < 10$. 
    We find no systematic offset between the two Galactic latitude groups, indicating that any discrepancies in label estimation primarily arise from the increased scatter caused by extinction. 
    The presence of extinction introduces an inherent latent variable that is intertwined within the spectra, which we neglected. 
    Unlike the derivation of stellar labels through direct forward modeling, it is crucial to consider the existence of extinction as an important parameter. 
    The \xp\ simplifies the model by choosing to ignore the presence of extinction, as it does not significantly influence our conclusions, as demonstrated in the validation process.
    We refer users whose scientific focus revolves around the extinction of \gaia\ XP to the work of \cite{2023arXiv230303420Z} and related references there.
    
    \item 
    We make the assumption that the training data, obtained from the overlap of \gaia\ and \apogee\ datasets, consitutes a representative and sufficiently diverse sample of stars.
    Our main focus is on deriving accurate stellar parameters and abundances for red giant branch (RGB) stars, which are well-covered by \apogee\ observations. 
    To ensure the purity of the RGB catalog, we exclude white dwarfs and hot stars from our training set. 
    Although this exclusion is expected to have minimal impact on the training process, there is a possibility that some hot stars may be inadvertently included in the RGB catalog if they fall within the predefined boundaries. 
    However, we have taken precautionary measures by examining the $G_{\rm BP} - G_{\rm RP}$ color index for the RGB catalog, and the fraction of misclassified hot stars is found to be negligible.

    \item 
     We impose a cutoff at \moh\ values larger than -2 for all stars in our analysis. 
     This decision is based on the limited coverage of the training sample of \apogee\ in terms of [M/H] values, which excludes very metal-poor stars. 
     The fraction of stars in the inner disk with \moh\ values below -2 is very small, approximately 0.003\% \citep{Rix2022}. 
     If the analysis requires a focus on very metal-poor stars with \moh$<-2$, we recommend referring to works such as \cite[]{lihaining2022, 2023arXiv230317676Y}, which specifically address the very metal-poor population.
    
    \item Our assumption of all sources being single stars overlooks the presence of a significant fraction of stars in binary systems. 
    While our model can accurately predict the labels for the primary star in binaries with large mass (and light) ratios, the performance may be impacted for systems with close to equal mass ratios. This limitation is inherent in many data-driven methods.
    A possible strategy to tackle this issue is to explicitly consider each spectrum as a binary system. 
    This approach involves comparing the goodness-of-fit, such as the reduced $\chi^2$, between single-star and binary solutions for each spectrum \citep{El-Badry2018}. 
    By doing so, it becomes feasible to identify potential binary systems (Niu et al. \textit{in preparation}). 
    This avenue holds promise for further refining our understanding of binaries within the XP sample.

    \item 
    % Some of the identified red clump (RC) stars with a \logg\ value around 2.5 exhibit larger uncertainties, which make their \moh\ and \aom\ values less reliable during the validation of $\alpha$-abundance in non-disk populations (see subsection~\ref{subsubsec:ges}). 
    % However, this issue is not found in disk stars, suggesting that it may be due to the absence of RC stars and stars below the luminosity of the RC in the \apogee\ training sample, specifically within the halo substructure and LMC.
    For metal-poor (\moh$<-1$) stars, the  \aom\ uncertainties are relatively higher for stars with higher surface gravity (\logg$>2.5$) compared to metal-rich stars. This is especially true for $\alpha$-poor stars, as observed in the validation process described in Subsection \ref{subsubsec:ges}.
    This could be attributed to the fact that high \logg\ values may result in pressure-broadened wings of strong metal lines (e.g., Mg I and Ca I, see \citealt{Gray2008} for details).
    In addition to line broadening effects, the inherent weak metal features in metal-poor stars can pose challenges in accurately estimating \aom, given the existing correlation between \aom\ and \moh.
    For studies focusing on metal-poor stars, we recommend selecting stars with \logg\ values lower than 2.5 and carefully considering the error estimation provided by \xp\ to ensure the accuracy and validity of the derived parameters.
\end{itemize}

\subsection{Conclusion}

This paper presents the \xp\ model, a data-driven approach for performing non-linear regressions that estimate stellar labels from the low-resolution spectroscopic data provided by the BP/RP (XP) spectra from \gaia\ DR3. Our approach has two new aspects compared to published analyses: it also yields precise estimates for \aom~, and it imploys the \emph{hallucinator}.
By utilizing a pre-trained model based on high-resolution \apogee\ spectra, \xp\ with the \emph{hallucinator} component, achieves remarkable accuracy in predicting the effective temperature, surface gravity, metallicity, and $\alpha$-abundance of stars. 
Through 2-fold cross-validation, the model demonstrates accuracies of approximately 50~K, 0.12~dex, 0.07~dex, and 0.02~dex, respectively.
Our study results in a comprehensive catalog containing fundamental parameters (\teff\ and \logg) and abundance prediction (\moh\ and \aom) for approximately 23 million RGB stars. 
This extensive dataset, accompanied by the open-source code, is publicly accessible at \url{https://zenodo.org/record/8002699}.

The extensive catalog of stellar labels for $\sim$23 million RGB stars generated in this study provides the astronomical community with a valuable multi-purpose data resource. 
The unprecedented scale of all-sky \aom measurements will facilitate novel insights into the formation history and chemodynamics of the Milky Way. 
Additionally, the public release of the open-source \xp\ code will promote further methodological advancements in analyzing low-resolution spectra. 
This work demonstrates the meaningful \aom~ abundance information that can be extracted from \gaia's low-resolution spectroscopic data. By developing innovative techniques tailored to these spectra, as with \xp, we can unleash the full potential of the vast XP dataset to illuminate our Galaxy's chemical evolution. 
The catalog and methods presented here will enable a diverse range of Galactic archaeology research and upcoming spectroscopic surveys.

% The \xp\ model signifies a remarkable way in the estimation of stellar parameters using low-resolution spectra, facilitating the characterization of a vast population of stars with exceptional precision. 
% The catalog generated by this study presents a valuable resource for numerous astronomical surveys, including but not limited to the upcoming LAMOST-III survey, as well as for subsequent high-resolution follow-up observations. 

%==============================================================

%\begin{acknowledgments}

\noindent
\vspace{10mm}

It is a pleasure to thank Chao Liu (NAOC), Haibo Yuan (BNU), Rene Andrae (MPIA), Dongwei Fan (NAOC), Bo Zhang (NAOC), Hao Tian (NAOC) for help with the project.
This project was developed in part at the 2023 Gaia XPloration, hosted by the Institute of Astronomy, Cambridge University.
% J.L. thanks the National Key R\&D Program of China No. 2019YFA0405500, and the National Natural Science Foundation of China (NSFC) with grant No. 11835057. 

This work has made use of data from the European Space Agency (ESA) mission {\it Gaia} (\url{https://www.cosmos.esa.int/gaia}), processed by the {\it Gaia} Data Processing and Analysis Consortium (DPAC, \url{https://www.cosmos.esa.int/web/gaia/dpac/consortium}). Funding for the DPAC has been provided by national institutions, in particular the institutions participating in the {\it Gaia} Multilateral Agreement.

Funding for the Sloan Digital Sky Survey IV and V has been provided by the Alfred P. Sloan Foundation, the U.S Department of Energy Office of Science, and the Participating Institutions. 
SDSS-IV acknowledges support and resources from the Center for High-Performance Computing at the University of Utah. 
The SDSS website is \url{www.sdss.org}.
SDSS-IV and SDSS-V have been managed by the Astrophysical Research Consortium for the Participating Institutions of the SDSS Collaboration including the Brazilian Participation Group, the Carnegie Institution for Science, Carnegie Mellon University, the Chilean Participation Group, the French Participation Group, Harvard-Smithsonian Center for Astrophysics, Instituto de Astrofísica de Canarias, The Johns Hopkins University, Kavli Institute for the Physics and Mathematics of the Universe/University of Tokyo, the Korean Participation Group, Lawrence Berkeley National Laboratory, Leibniz Institut für Astrophysik Potsdam, Max-Planck-Institut für Astronomie (Heidelberg), Max-Planck-Institut für Astrophysik (Garching), Max-Planck-Institut für Extraterrestrische Physik, National Astronomical Observatories of China, New Mexico State University, New York University, University of Notre Dame, Observatário Nacional/MCTI, The Ohio State University, Pennsylvania State University, Shanghai Astronomical Observatory, United Kingdom Participation Group, Universidad Nacional Autónoma de México, University of Arizona, University of Colorado Boulder, University of Oxford, University of Portsmouth, University of Utah, University of Virginia, University of Washington, University of Wisconsin, Vanderbilt University, and Yale University.

Guoshoujing Telescope (the Large Sky Area Multi-Object Fiber Spectroscopic Telescope LAMOST) is a National Major Scientific Project built by the Chinese Academy of Sciences. Funding for the project has been provided by the National Development and Reform Commission. LAMOST is operated and managed by the National Astronomical Observatories, Chinese Academy of Sciences.

This work made use of the Third Data Release of the GALAH Survey. The GALAH Survey is based on data acquired through the Australian Astronomical Observatory, under programs: A/2013B/13 (The GALAH pilot survey); A/2014A/25, A/2015A/19, A2017A/18 (The GALAH survey phase 1); A2018A/18 (Open clusters with HERMES); A2019A/1 (Hierarchical star formation in Ori OB1); A2019A/15 (The GALAH survey phase 2); A/2015B/19, A/2016A/22, A/2016B/10, A/2017B/16, A/2018B/15 (The HERMES-TESS program); and A/2015A/3, A/2015B/1, A/2015B/19, A/2016A/22, A/2016B/12, A/2017A/14 (The HERMES K2-follow-up program). We acknowledge the traditional owners of the land on which the AAT stands, the Gamilaraay people, and pay our respects to elders past and present. This paper includes data that has been provided by AAO Data Central (datacentral.org.au).

This work has made use of the Python package {\tt\string GaiaXPy}, developed and maintained by members of the Gaia Data Processing and Analysis Consortium (DPAC), and in particular, Coordination Unit 5 (CU5), and the Data Processing Centre located at the Institute of Astronomy, Cambridge, UK (DPCI).

% This work made use of the Third Data Release of the GALAH Survey. The GALAH Survey is based on data acquired through the Australian Astronomical Observatory, under programs: A/2013B/13 (The GALAH pilot survey); A/2014A/25, A/2015A/19, A2017A/18 (The GALAH survey phase 1); A2018A/18 (Open clusters with HER- MES); A2019A/1 (Hierarchical star formation in Ori OB1); A2019A/15 (The GALAH survey phase 2); A/2015B/19, A/2016A/22, A/2016B/10, A/2017B/16, A/2018B/15 (The HERMES-TESS program); and A/2015A/3, A/2015B/1, A/2015B/19, A/2016A/22, A/2016B/12, A/2017A/14 (The HERMES K2-follow-up program). We acknowledge the traditional owners of the land on which the AAT stands, the Gamilaraay people, and pay our respects to elders past and present. This paper includes data that has been provided by AAO Data Central (datacentral.org.au).

% This work has made use of the Python package {\tt\string GaiaXPy}, developed and maintained by members of the Gaia Data Processing and Analysis Consortium (DPAC), and in particular, Coordination Unit 5 (CU5), and the Data Processing Centre located at the Institute of Astronomy, Cambridge, UK (DPCI).
%  \end{acknowledgments}

  \clearpage
  \vspace{5mm}
  \facilities{\textsl{Gaia}, \textsl{APOGEE}}

  \software{
  {\tt\string PyTorch} \citep{NEURIPS2019_9015},
  {\tt\string Astropy} \citep{2018AJ....156..123A}, 
  {\tt\string Scipy} \citep{2020SciPy-NMeth}, 
  {\tt\string scikit-learn} \citep{scikit-learn}, 
  TOPCAT \citep{2005ASPC..347...29T}, 
  {\tt\string smplotlib} \citep{jiaxuan_li_2023_7847258}.
  }

\appendix

\section{Query avaiable XP data}\label{app:220m}
\begin{lstlisting}
select * from gaiadr3.gaia_source 
where has_xp_continuous='true'
\end{lstlisting}

\section{Validation by disk stars} \label{app:disk}

\begin{figure*}[htbp]
    \includegraphics[width=\linewidth]{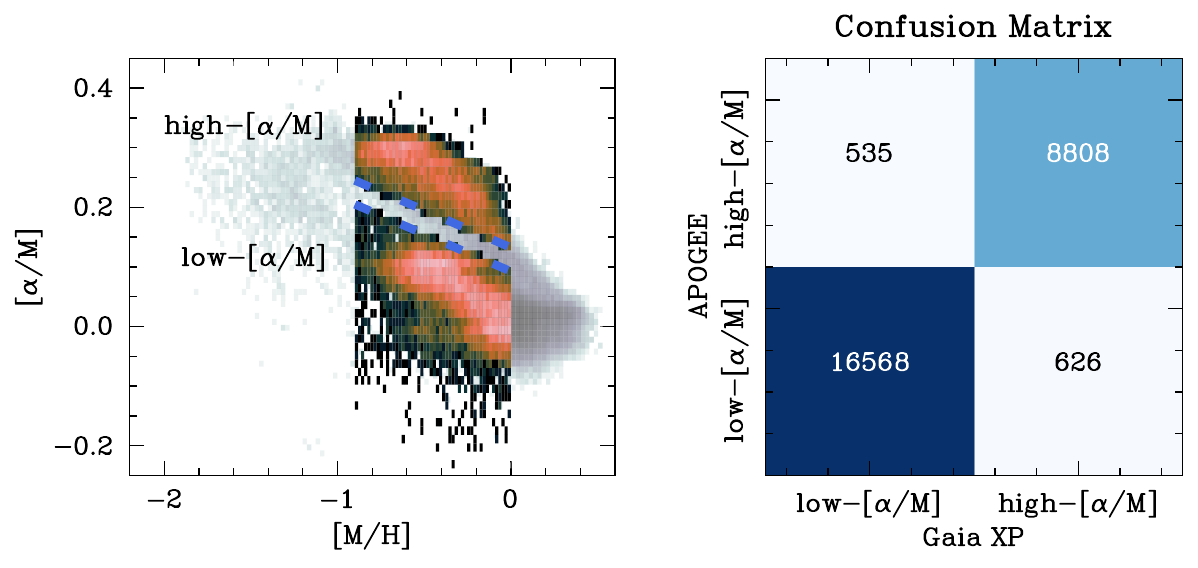}
    \caption{
    Left: Selected stars classified into low-\aom\ and high-\aom\ groups with $-0.9<$\moh$<0$, representing the ground truth from \apogee.
    Right: Confusion matrix showing the \gaia\ XP-predicted low/high \aom\ group of \apogee\ group.
    }
    \label{fig:disk_alpha}
\end{figure*}

We conduct an analysis to evaluate the discriminative power of \xp\ in distinguishing between different chemical components of the disk, namely the low-alpha and high-alpha disk, while considering stars within the same range of metallicity. 
By comparing the \xp\ classifications with the given APOGEE labels, we can assess how effectively \xp\ can differentiate between these two disk components based on their \aom.
To select high-alpha and low-alpha disk stars based on APOGEE labels, we use a specific metallicity range ($-0.9 <$ [M/H] $< 0$) as shown in Fig~\ref{fig:disk_alpha}. 
All of the selected stars belonging to the low and high disk groups, as classified by \apogee, have undergone 2-fold cross-validated estimations of \moh\ and \aom\ using \xp.
To evaluate the performance of \xp\ in distinguishing between high-alpha and low-alpha disk stars, we employ precision and recall metrics. 
Precision measures the proportion of correctly classified stars out of all stars classified as a particular category, while recall measures the proportion of correctly classified stars out of all stars belonging to that category.

As illustrated in Fig~\ref{fig:disk_alpha}, our validation results highlight the remarkable performance of \xp\ in differentiating between the low-alpha and high-alpha disk populations. For the low-alpha disk stars, 97\% of the stars identified by \xp\ correspond to the low-alpha class as defined by \apogee\ labels. Furthermore, \xp\ accurately classifies 96\% of the entire low-alpha disk sample identified by \apogee. Similarly, for the high-alpha disk, 93\% (8,808 out of 9,434) of the stars identified by \xp\ represent the high-alpha class, and 94\% (8,808 out of 9,343) of the complete high-alpha sample identified by \apogee\ are correctly recognized by \xp.

\section{Validation by different Galactic latitudes}\label{app:further_validation}

\begin{figure*}[htbp]
    \centering
    \includegraphics[width=\linewidth]{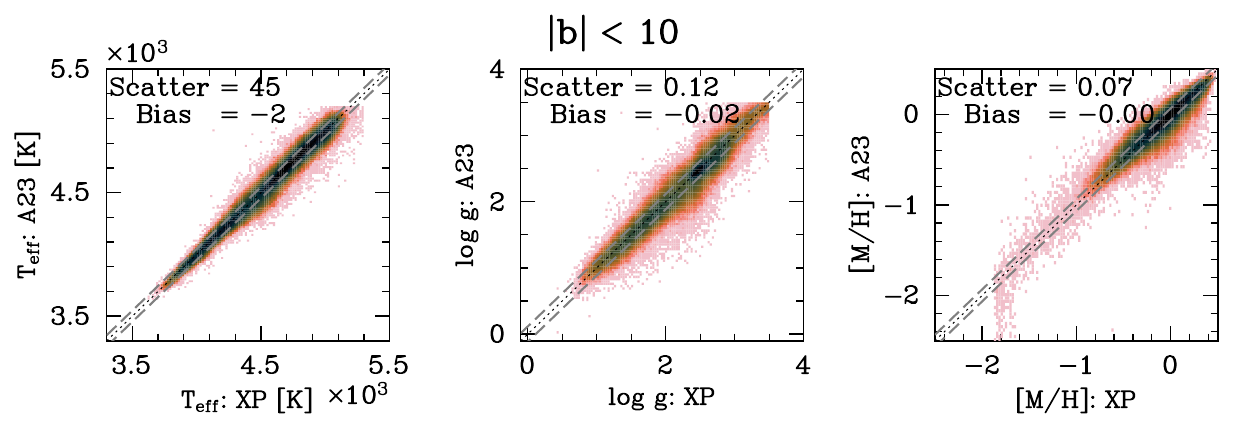}
    \includegraphics[width=\linewidth]{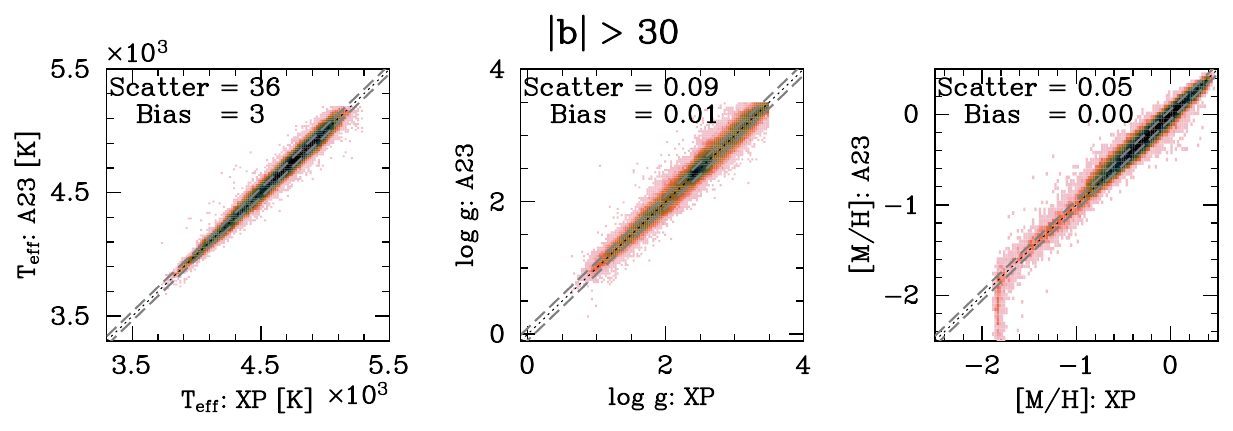}
    \caption{
    The comparison of the \teff, \logg, and \moh\ parameters, divided into different Galactic latitudes ($b$), from left to right panel. 
    The top panel corresponds to stars with $\mid b \mid < 10$, while the bottom panel represents stars with $\mid b \mid > 30$. 
    The comparison is made using the cross-matched catalog with \citetalias{2023arXiv230202611A}'s RGB sample \citep{2023arXiv230202611A}.
    }
    \label{app:C}
\end{figure*}

We present a detailed comparison of the \teff, \logg\ and \moh\ divided into two different Galactic latitudes ($b$) ($| b | < 10$ and $| b | > 30$) as illustrated by Fig.\ref{app:C}.

\bibliography{main}{}
\bibliographystyle{aasjournal}



\end{document}